\pdfoutput=1 

\documentclass[letterpaper,11pt]{article}
\usepackage{tabularx} 
\usepackage{amsmath}  
\usepackage{amssymb}  
\usepackage{graphicx} 
\usepackage[margin=1in,letterpaper]{geometry} 
\usepackage[final]{hyperref} 
\hypersetup{
	colorlinks=false,       
	linkcolor=blue,        
	citecolor=blue,        
	filecolor=magenta,     
	urlcolor=blue         
}

\usepackage[affil-it]{authblk}  

\usepackage[font=footnotesize,labelfont=bf]{caption} 

\usepackage{physics} 
\usepackage{slashed} 

\usepackage{tikz-feynman} 
\usepackage{xcolor} 
\usepackage{xfrac}
\usepackage{float} 
\usepackage{tabularray} 

\usepackage{subcaption}
\usepackage{wrapfig}

\usepackage{bbold}
\usepackage{bm}
\usepackage[normalem]{ulem}

\usepackage{titlesec}
\titleformat*{\subsection}{\bfseries\boldmath}

\renewcommand{\L}{\mathcal{L}}

\newcommand{\M}{\mathcal{M}}
\renewcommand{\O}{\mathcal{O}}

\newcommand{\PL}{\text{P}_\text{L}}
\newcommand{\PR}{\text{P}_\text{R}}

\newcommand{\MSbar}{$\overline{\text{MS}}$~}

\newcommand{\lambar}{\bar\lambda}
\newcommand{\re}{\widetilde\Re}
\newcommand{\ddmu}[1]{\mu\frac{d #1}{d\mu}}
\newcommand{\ppmu}[1]{\mu\frac{\partial #1}{\partial\mu}}

\usepackage[style=ieee, citestyle=numeric-comp, backend=biber, sorting=none]{biblatex}
\title{\bfseries\boldmath On-shell Renormalization with Vector-like Leptons,
  One-loop Muon--Higgs Coupling and Muon $g-2$}

\author[1]{Kilian Möhling \thanks{kilian.moehling@tu-dresden.de}}
\author[1]{Dominik Stöckinger \thanks{dominik.stoeckinger@tu-dresden.de}} 
\author[1]{Hyejung Stöckinger-Kim \thanks{hyejung.stöckinger-kim@tu-dresden.de}} 
\affil[1]{Institut für Kern- und Teilchenphysik, TU Dresden, Zellescher Weg 19, 01069 Dresden, Germany}
\date{}

\begin{document}
\maketitle


\begin{abstract}
        Models with vector-like leptons can strongly modify the lepton
        mass generation mechanism and lead to correlated effects in
        lepton--Higgs couplings and lepton dipole moments. Here we
        begin an analysis of higher-order corrections in such models
        by setting up a renormalization scheme with full on-shell
        conditions on the lepton self energies, masses and fields. A
        minimal set of fundamental parameters is renormalized in the
        \MSbar scheme.
        We provide a detailed discussion of
        lepton mixing and redundancies at higher orders, show how the relevant counterterms
	can be obtained from the renormalization conditions, and
        determine the $\beta$-functions corresponding to the scheme. 
	As a first application we calculate the one-loop effective muon--Higgs coupling and analyse its correlation with the muon anomalous magnetic moment $\Delta a_\mu^{\text{VLL}}$. In the interesting case of large masses and
        opposite-sign coupling, the lowest-order correlation implies a
        fixed value of $\Delta a_\mu^{\text{VLL}}$ around $22.5\times 10^{-10}$,
        while the higher-order corrections significantly reduce this
        value to the interval $(10\ldots18)\times 10^{-10}$.
\end{abstract}

\setcounter{tocdepth}{2}
\tableofcontents

\section{Introduction}

In recent years vector-like lepton (VLL) models have gained increasing interest as a potential answer to some of the long standing open
questions in particle physics. 
We focus on a model where the Standard Model (SM) is extended by
additional vector-like leptons which can have  fundamental
gauge-invariant  Dirac masses and mix with the SM leptons.
Such VLLs have been considered in the context of the muon $g-2$ (also in combination with extended scalar sectors)
\cite{Kannike:2011ng,Dermisek:2013gta,Poh:2017tfo,Dermisek:2015oja,Chun:2020uzw,Dermisek:2020cod,Frank:2020smf,Dermisek:2021ajd,Lee:2022nqz,Dermisek:2023tgq}, 
charged lepton flavor violation (cLFV) \cite{Ishiwata:2013gma,Falkowski:2013jya,Poh:2017tfo}, 
and electroweak precision observables \cite{Kannike:2011ng,Dermisek:2013gta,Ellis:2014dza}.

The key feature of such VLL models lies in their change of the lepton
 	mass generation mechanism: The masses of the SM-like leptons result in part (or even entirely) from mixing, rather than through their
 	fundamental interactions with the Higgs only.
In general, SM fermion mass terms require a source of chirality flips and a breaking of electroweak gauge invariance. 
These two requirements are shared by other observables such as effective Higgs-lepton couplings $\lambda _{l l'}$ 
and lepton anomalous magnetic moments $a_l = (g-2)_l / 2$. 
The implications of these relations for $a_\mu$ are discussed in Refs.~\cite{Athron:2021iuf,Stockinger:2022ata}
as well as in Refs.~\cite{Dermisek:2013gta,Crivellin:2021rbq}
with focus on the correlation between $a_\mu$, $\lambda_{\mu\mu}$ and further observables in a wide range of models.
Many extensively studied models such as two-Higgs-doublet (2HDM), supersymmetric (SUSY) and leptoquark (LQ) models
introduce modifications in the scalar sector, which lead to new chirality flips at the loop level. 
In contrast, VLL models provide new sources of chirality flips already at tree-level via the Dirac masses
and Yukawa couplings of the VLL fields and imply distinct correlations between the observables involving chirality flips \cite{Dermisek:2013gta,Dermisek:2023tgq,Dermisek:2023nhe}.

These correlations can be understood as a result of effective dimension-6 operators of the form $\bar{l} l HH ^\dagger H$ that 
are generated at tree-level in VLL models. Such operators 
can lead to $\O(1)$ tree-level corrections to $\lambda_{\mu\mu}$ that turn out to be strongly correlated with 
contributions to $a_\mu$ (for a general study of these operators see Ref.~\cite{Goudelis:2011un}). 
On the other hand, experimental constrains restrict $|\lambda_{\mu\mu}|$ to be close to the SM value. 
This allows either $\lambda_{\mu\mu} \approx \lambda_{\mu\mu}^\text{SM}$ in which case $\Delta a_\mu^\text{VLL} \approx 0$ or,
intriguingly, $\lambda_{\mu\mu} \approx -\lambda_{\mu\mu}^\text{SM}$ in which case the correlation implies some fixed
value $\Delta a_\mu^\text{VLL} \neq 0$. 
For example, in the case of one generation of heavy VLLs coupling to the muon discussed in Refs.~\cite{Kannike:2011ng,Dermisek:2013gta}, the tree-level correlation predicts 
\begin{align}
\label{eq:amuVLL}
a_\mu ^{\text{VLL}}|_{(\lambda_{\mu\mu} = - \lambda_{\mu\mu} ^{\text{SM}})} &\simeq 22.5 \times 10 ^{-10}\,.
\end{align}
This is intriguingly close to the deviation between the Fermilab Run 1,2,3 measurements \cite{Muong-2:2021ojo,Muong-2:2023cdq} 
and the 2020 white paper SM theory value based on \cite{Aoyama:2020ynm,Davier:2017zfy,Keshavarzi:2018mgv,Colangelo:2018mtw,Hoferichter:2019mqg,
Davier:2019can,Keshavarzi:2019abf,Kurz:2014wya,Melnikov:2003xd,Masjuan:2017tvw,Colangelo:2017fiz,Hoferichter:2018kwz,Gerardin:2019vio,
Bijnens:2019ghy,Colangelo:2019uex,Colangelo:2014qya,Blum:2019ugy,Aoyama:2012wk,Aoyama:2019ryr,Czarnecki:2002nt,Gnendiger:2013pva}, 
$a_\mu^\text{exp} - a_\mu^\text{SM} = (24.9\pm 4.8)\times
10^{-10}$. Clearly, scrutiny of SM predictions (see
Refs.~\cite{Aoyama:2020ynm,Colangelo:2022jxc} for an overview of
results and prospects) and Fermilab Run 4,5,6 results might change this value. Hence it is an open question whether the deviation in Eq.~\eqref{eq:amuVLL} remains a viable value.

The aim of the present paper is to provide a starting point for higher-order studies of VLL models. 
Here, we focus on the SM extended by one generation of VLLs that mix with all of the SM leptons.
We develop an on-shell renormalization scheme and discuss the treatment of the lepton mixing at higher orders. 
As a first application of the scheme, we determine the modification of Eq.~\eqref{eq:amuVLL} 
from including one-loop corrections to the effective muon-Higgs coupling $\lambda_{\mu\mu}$. 

On-shell renormalization has already been studied in a variety of scenarios beyond the Standard Model (BSM)
and has often turned out to be highly nontrivial. This is in
contrast to the electroweak SM where there is a one-to-one
correspondence between fundamental parameters and masses or similarly
simple on-shell observables. E.g. the on-shell renormalization of the Minimal Supersymmetric Standard Model
(MSSM) has been studied in Refs.~\cite{Hollik:2002mv,Fritzsche:2002bi,Hollik:2003jj} with focus
on the SUSY sectors; here, SUSY
relations between masses imply that only a subset of masses can be on-shell renormalized. 
In addition, the renormalization of parameters such as $\tan\beta$ or
mixing angles is required and troublesome \cite{Freitas:2002um,Heinemeyer:2010mm}.
The on-shell renormalization of the two-Higgs doublet model (2HDM) has
been studied in
Refs.~\cite{Santos:1996vt,Kanemura:2004mg,Krause:2016oke,Kanemura:2017wtm,Altenkamp:2017ldc,Denner:2018opp,Grimus:2018rte}
with increasing focus on mixing angles and similar parameters for
which an on-shell scheme is not suitable. For these parameters
\MSbar renormalization conditions can be advantageous,
but in such mixed schemes the renormalization of the
vacuum expectation values/of tadpoles becomes 
critical~\cite{Freitas:2002um,Altenkamp:2017ldc,Krause:2016oke,Denner:2018opp,Grimus:2018rte,Denner:2019vbn,Dittmaier:2022maf,Dittmaier:2022ivi,Dudenas:2020ggt}.

In the VLL model considered in this paper, the $5\times5$ lepton mixing is responsible for the intriguing 
phenomenology, but it also complicates the discussion of higher-order corrections. 
Nonetheless, with the steady increase in experimental precision as well as spectacular predictions like Eq.\
(\ref{eq:amuVLL}), increasing the theoretical accuracy is desirable. This requires a proper definition of a renormalization
scheme, where an on-shell scheme allows to study correlations between observables in a particularly transparent way.

We begin with a concise review of the model in Sec.~\ref{sec:model}, including the definition, the conventions for the Dirac masses, the Yukawa interaction parameters, and the unitary mixing matrix for lepton mass eigenstates.  
In Sec.~\ref{sec:scheme} we develop a renormalization scheme for this
model where all masses and fields are on-shell renormalized and additional 
$\overline{\text{MS}}$-conditions are imposed on original parameters of the gauge-invariant Lagrangian. We discuss in detail the counterterm structure 
and solution to the on-shell condition as well as the residual renormalization scale dependence of the fundamental parameters.
Afterwards, in Sec.~\ref{sec:one-loop} we extend the definition of the
effective muon-Higgs coupling to one-loop order and apply the renormalization scheme to compute
the corresponding renormalized higher-order contributions. Finally, in Sec.~\ref{sec:conclusion} we summarize the results and our conclusion.

\section{Model Definition}\label{sec:model}
\begin{table}[h]
	\centering
	\renewcommand{\arraystretch}{1.25}%
	\begin{tabular}{|c|c|c|}
		\hline
		& SM & VL \\ \hline\hline
		& $l_L^i = \begin{pmatrix} \nu^i_L \\ e^i_L \end{pmatrix} ~~~ e_R^i ~~~ H = \begin{pmatrix} \phi^+ \\ v + \frac{1}{\sqrt{2}} (h + i\phi^0) \end{pmatrix}$ & 
		$L_{L/R} = \begin{pmatrix} L^0_{L/R} \\ L^-_{L/R} \end{pmatrix} ~~~ E_{L/R}$\\ \hline
		$\text{SU}(2)_\text{L}$ \vphantom{\Big|} & \textbf{2} \hspace{1cm} \textbf{1} \hspace{2.5cm} \textbf{2} \hspace{.5cm} & \hspace{.5cm} \textbf{2} \hspace{2cm} \textbf{1} \\ \hline
		$\text{U}(1)_\text{Y}$ \vphantom{\Big|} & $-\frac{1}{2}$ \hspace{.8cm} $-1$ \hspace{2.4cm} $\frac{1}{2}$ \hspace{.75cm} & \hspace{.25cm} $ -\frac{1}{2}$ \hspace{1.75cm} $-1 $ \\ \hline
	\end{tabular}
	\caption{Definition of the SM and VL lepton fields (with
          generation index $i=1,2,3$) and their corresponding quantum
          numbers (where the electric charge is related to SU(2)$_L$
          and U(1)$_Y$ quantum numbers as $Q=T^3+Y$). The fields $h,~\phi^0,~\phi^+$ correspond to 
		the physical Higgs field and the neutral and charged
                Goldstone bosons, and $v\approx 174$ GeV. The VLL
                fields are defined as $L_{L/R} = P_{L/R}L$ and
                $E_{L/R} = P_{L/R}E$, where $L$ are $SU(2)$ doublets
                and  $E$ $SU(2)$ singlets with the same quantum numbers as the SM lepton doublets and singlets respectively.}
	\label{tab:leptons}
\end{table}
The SM lepton sector consists of three generations of left-handed $\text{SU(2)}_L$ doublets $l_L^i$ and right-handed singlets $e_R^i$.
In addition, we consider one generation of vector-like leptons (VLL) comprised of both left- and right handed doublets $L_{L/R}$ and singlets $E_{L/R}$ with gauge-quantum numbers
identical to those of the SM fields (see Table \ref{tab:leptons}). This model has previously been studied in the muon-specific limit in Refs. \cite{Kannike:2011ng,Dermisek:2013gta}
and in the context of charged lepton flavor violation in Ref. \cite{Poh:2017tfo}.
It is useful to combine the SM and VL lepton fields with identical quantum numbers into the 4-tuples
\begin{align}
	l^a_L = (l^i_L, L_L)^{\text{T}}, \qquad e^a_R = (e^i_R,E_R)^{\text{T}} \qquad \text{for} \qquad a = 1,2,3,4.
\end{align}
Viewed in this way, the model contains four generations of the usual chiral leptons $l^a_L$ and $e^a_R$ as well as one generation of opposite-chirality fields $L_R$ and $E_L$.
Using these definitions, the most general renormalizable Lagrangian can be written as
\begin{align}\label{eq:L-sym}
	\begin{split}
		\L &\supset \bar{l}^a_L i\slashed{D} l^a_L + \bar{e}^a_R i\slashed{D} e^a_R + \bar L_R i\slashed{D}L_R + \bar E_L i\slashed{D} E_L  \\
		&\quad - \big(\bar{l}^a_L M_L^a L_R + \bar E_L M_E^a e^a_R + \bar{l}^a_L Y_{ab} e^b_R H + \bar E_L \bar\lambda L_R H^\dagger  + h.c. \big),
	\end{split}
\end{align}
where the $\text{SU}(2)_\text{L}\times\text{U}(1)_\text{Y}$ covariant
derivative is given by $D_\mu = \partial_\mu - i g_W
\frac{\sigma^i}{2} W^i_\mu + i g_Y Y B_\mu$. In the present paper we
assume CP invariance and consider only real parameters.
$M_L$ and $M_E$ are 4-tuple vectors of mass parameters, $Y$ is a
$4 \times 4$ Yukawa matrix and $\lambar$ a single Yukawa coupling.
Similar to the SM Yukawa sector, the Lagrangian Eq.~\eqref{eq:L-sym} contains a redundancy and not all parameters carry physical information.
This redundancy results from the invariance of the Lagrangian under the following simultaneous transformation of the fields and parameters
\begin{align}\label{eq:wlog-transformation}
	l_L^a \to (V_L l_L)^a, \quad e^a_R \to (V_R e_R)^a, \quad  M^a_L \to (V_L M_L)^a, \quad M_E^a \to (M_E^{\text{T}} V_R^\dagger)^a, \quad Y_{ab} \to (V_L Y V_R^\dagger)_{ab},
\end{align}
where $V_{L/R}$ are arbitrary unitary $4\times 4$ matrices. As a
result, 12 of the Yukawa and mass parameters can be absorbed into a redefinition of the lepton fields.
In particular, it is convenient to absorb the off-diagonal Dirac
masses together with the off-diagonal SM Yukawa couplings, such that
w.l.o.g. the mass and coupling matrices can be written as
\begin{align}
  \label{eq:parametrization-symm-pars}
  M^a_L=(0,0,0,m_L)^{\text{T}}, \qquad M^a_E = (0,0,0,m_E)^{\text{T}}, \qquad Y_{ab} = \begin{pmatrix} y_i \delta_{ij} & \lambda^E_i \\ \lambda^L_j & \lambda \end{pmatrix}.
\end{align}
The remaining 12 physical parameters in this sector are thus taken as
the 3 SM Yukawa couplings $y_i$, the 6 new Yukawa couplings
$\lambda^L_i,\lambda^E_i$ which are responsible for the interactions
between the SM and VLL fields, the Yukawa coupling $\lambda$ involving the VLL fields,
and the 2 Dirac masses $m_L$ and $m_E$. We note that the 13th new physical parameter $\lambar$ cannot be absorbed by any field redefinition.
The vacuum expectation value (vev) of the Higgs field spontaneously breaks the $\text{SU}(2)_\text{L}\times\text{U}(1)_\text{Y}$ symmetry and 
generates additional lepton mass terms in the Lagrangian. 
After electroweak symmetry breaking (EWSB), the $5\times 5$ mass matrices for the charged leptons and neutrinos are given by
\begin{align}\label{eq:lepton-mass-matrix}
	(\bar e_L^a, \bar E_L) \begin{pmatrix} (Yv)_{ab} & M_L^a \\ M_E^{b} & \lambar v \end{pmatrix} \begin{pmatrix} e_R^b \\ L^-_R \end{pmatrix}, \qquad
	(\bar\nu_L^a,0) \begin{pmatrix} 0 & M_L^a \\ 0 & 0 \end{pmatrix} \begin{pmatrix} 0 \\ L^0_R \end{pmatrix}.
\end{align}
We introduce the lepton mass eigenstate fields as the 5-tuples
\begin{equation}\label{eq:mass-basis}
		\hat e_L = U^\dagger_L \begin{pmatrix}	e^a_L \\ E_L \end{pmatrix}, \qquad \hat e_R = U^\dagger_R \begin{pmatrix}	e^a_R \\ L^-_R \end{pmatrix}, \qquad
		\hat\nu_L = U_L^{\nu\dagger} \begin{pmatrix}\nu^a_L \\ 0 \end{pmatrix}, \qquad \hat\nu_R = U_R^{\nu\dagger}\begin{pmatrix}	0 \\ L^0_R \end{pmatrix},
\end{equation}
where the unitary $5\times 5$ matrices $U_{L/R}$ and $U^\nu_{L/R}$ are chosen such that the mass matrices are diagonalized as
\begin{align}\label{eq:lepton-mass-diagonalization}
	\left[U_{L}^\dagger \begin{pmatrix} Y v & M_L \\ M_E^{\text{T}} & \lambar v \end{pmatrix} U_{R}\right]_{ab} := m_a \delta_{ab}, \qquad
	\left[U_{L}^{\nu\dagger}\begin{pmatrix} 0 & M_L \\ 0 & 0 \end{pmatrix} U^\nu_{R}\right]_{ab} := m_L \delta_{a4}\delta_{b4}
\end{align}
Note that in our convention the only massive neutrino is $\hat\nu^4$ with mass $m_L$, while the SM neutrinos $\hat\nu_i$ remain massless and $\hat\nu_5=0$.
The masses of the charged leptons are given approximately by
\begin{subequations}
	\begin{align}
		m_i &= y_i v + \frac{\lambar\lambda^E_i \lambda^L_i v^3}{m_E m_L} + \O(v^4)\\
		m_{4,5} &= m_{L,E} + \O(v^2)\,,
	\end{align}
\end{subequations}
where $y_i$ is defined in Eq.~(\ref{eq:parametrization-symm-pars}).  
In the mass eigenstate basis the lepton mass terms and physical Higgs-lepton Yukawa interactions are given by the following terms in the classical Lagrangian
\begin{align}\label{eq:L-SSB}
	\begin{split}
		\L &\supset - m_a \bar{\hat e}^a_L \hat e^a_R - m_L \bar{\hat\nu}^4_L \hat\nu^4_R - \tfrac{1}{\sqrt{2}} \bar{\hat e}^a_L Y^h_{ab} \hat e_R^b h + h.c.,
	\end{split}
\end{align} 
where we have introduced the effective $5\times 5$ Yukawa coupling matrix
\begin{align}
	Y^h = U_{L}^\dagger \begin{pmatrix} Y & 0 \\ 0 & \lambar \end{pmatrix} U_{R}\, .
\end{align}
Our convention differs slightly from the one used in Refs.\ \cite{Kannike:2011ng,Dermisek:2013gta} where $U^{(\nu)}_{L/R}=\mathbb{1}$ in the limit $v\to 0$, but which results in less symmetric expressions for the coupling and counterterm matrices.
All coupling matrices together with the Feynman rules are listed in appendix \ref{sec:Feynman-Rules}.

\section{Renormalization}\label{sec:scheme}
In this section we present a renormalization scheme developed for the
VLL model with full on-shell conditions on the lepton, Higgs- and gauge-boson self-energies. 
These on-shell conditions allow for a transparent computation of higher-order corrections to
observables in terms of the physical pole masses and well-defined
input parameters of the VLL model which satisfy
Eq.~\eqref{eq:lepton-mass-diagonalization} at all orders. Typically,
difficulties arise when the on-shell scheme is applied to the
renormalization of BSM scenarios since there is often no one-to-one
correspondence between masses and parameters. 
For example in the chargino and neutralino sector of the MSSM a simultaneous on-shell renormalization of all masses is impossible, as 3 of the 6 masses 
are fixed by the remaining input parameters \cite{Fritzsche:2002bi}. In the 2HDM a complete on-shell renormalization is possible,
but is made difficult by the appearance of additional parameters such as mixing angles and several possible schemes have been studied extensively in Refs.
 \cite{Santos:1996vt,Altenkamp:2017ldc,Krause:2016oke,Denner:2018opp,Grimus:2018rte}.
Our approach to the renormalization of the VLL model is similar to the treatment of the chargino and neutralino sector discussed in
Refs. \cite{Hollik:2002mv,Fritzsche:2002bi}. The important difference is that in our case a complete on-shell renormalization of all
lepton masses is possible.

\subsection{Gauge and Higgs Sector}\label{sec:gauge-higgs-scheme}
Here, we briefly list our notation for the one-loop on-shell (OS)
renormalization of the gauge and Higgs sector. It is chosen to be
identical to the one of the SM; hence we adopt the conventions and
results from Refs.~\cite{Denner-Böhm:2001,Denner:2019vbn}. The
renormalization transformation of the bare fields and parameters is done
on the level of mass-eigenstate quantities,
\begin{subequations}
	\begin{alignat}{3}
		&\quad\, h &&\to \big(1 + \tfrac{1}{2} \delta Z_h\big) h, \qquad &&M_{h}^2 \to M_h^2 + \delta M_h^2, \\
		&\begin{pmatrix}	A_\mu \\ Z_\mu \end{pmatrix} &&\to \begin{pmatrix}	1 + \tfrac{1}{2}\delta Z_{AA} & \tfrac{1}{2}\delta Z_{AZ} \\ \tfrac{1}{2}\delta Z_{ZA} & 1 + \tfrac{1}{2}\delta Z_{ZZ} 
		\end{pmatrix} \begin{pmatrix} A_\mu \\ Z_\mu \end{pmatrix}, \qquad && M^2_{Z} \to M_Z^2 + \delta M_Z^2,\\
		&~\, W^\pm_{\mu} &&\to \big(1 + \tfrac{1}{2} \delta Z_{WW}\big) W^\pm_\mu, && M_{W}^2 \to M_W^2 + \delta M_W^2,
	\end{alignat}
\end{subequations}
in terms of the renormalized quantities and renormalization constants
$\delta Z_h$, $\delta Z_{VV'}$ (with $V = A,Z,W$) as well as $\delta
M_h$, $\delta M_Z$ and $\delta M_W$. Denoting the unrenormalized Higgs and gauge-boson self-energies by
\begin{align}
	\Sigma_h(p^2), \qquad \Sigma^{\mu\nu}_{VV'}(q) = \Sigma^T_{VV'}(q^2) \big(\eta^{\mu\nu} - \tfrac{q^\mu q^\nu}{q^2}\big) + \Sigma^L_{VV'}(q^2) \tfrac{q^\mu q^\nu}{q^2},
\end{align}
the OS conditions on the renormalized Higgs and gauge-boson
self-energies result in the following expressions for the
renormalization constants,
\begin{subequations}\label{eq:boson-cts}
	\begin{alignat}{4}
		&\delta Z_h &&= - \re \tfrac{d}{dp^2}\Sigma_h(p^2)\big|_{p^2=M_h^2}, \qquad &&\delta M_h^2  &&= \re\Sigma_h(M_h^2),  \\
		&\delta Z_{VV} &&= - \re \tfrac{d}{dq^2}\Sigma^{T}_{VV}(q^2)\big|_{p^2=M_V^2}, \qquad &&\delta M_V^2 &&= \re \Sigma^{T}_{VV}(M_V^2), \\
		&\delta Z_{AZ} &&= -\tfrac{2}{M_Z^2} \re \Sigma^{T}_{AZ}(M_Z^2), \qquad &&\delta Z_{ZA} &&= \tfrac{2}{M_Z^2} \re \Sigma^{T}_{AZ}(0),
	\end{alignat}
\end{subequations}
where $\re$ removes the imaginary parts of any appearing loop integrals.
From charge universality it follows that the renormalization of the
electric charge is given by
\begin{align}
  e \to 
  e + \delta e, \qquad \tfrac{\delta e}{e} = - \tfrac{1}{2} \big(\delta Z_{AA} + \tfrac{s_W}{c_W}\delta Z_{ZA}\big).
\end{align}
In this setup the original parameters of the Lagrangian, i.e.\ in particular the Higgs vev $v$, are treated as a combination of the boson masses
and electric charge $e$. Consequently, the corresponding renormalization constant is fixed by the above counterterms as
\begin{align}\label{eq:dv}
	v \to v + \delta v, \qquad \tfrac{\delta v}{v} = \tfrac{\delta M_W^2}{2M_W^2} + \tfrac{c_W^2}{s_W^2} \Big(\tfrac{\delta M_Z^2}{2M_Z^2}-\tfrac{\delta M_W^2}{2M_W^2}\Big) - \tfrac{\delta e}{e}.
\end{align}
Finally, we discuss the renormalization of
tadpole diagrams and the implication for the renormalization of $v$. All previous equations and relations hold independently of the specific choice for the
tadpole counterterm $\delta t$, provided that the contribution of 1-particle reducible tadpoles $T$ and the tadpole counterterm to the self-energies in Eq.~\eqref{eq:boson-cts} are included.
However, different definitions of $\delta t$ lead to different values of the on-shell counterterms and subsequently $\delta v$. While this is inconsequential in the pure SM,
this choice can become important in BSM scenarios where not all parameters are necessarily renormalized on-shell. A discussion of this issue for the MSSM can be found in Ref. \cite{Freitas:2002um}
and further details along with several different schemes are given in Refs. \cite{Altenkamp:2017ldc,Denner:2018opp,Dudenas:2020ggt,Dittmaier:2022maf}.
Here we choose the parameter-renormalized tadpole scheme (PRTS) in
which the renormalized tadpole diagrams $T$ are set to zero,
\begin{align}\label{eq:tadpoleDef}
	T + \delta t^\text{PRTS} = 0.
\end{align}
With this requirement $v$ corresponds to the true minimum of the loop-corrected effective potential, but $\delta v$ is gauge-dependent. 
This scheme was found particularly advantageous in Ref.~\cite{Freitas:2002um} in conjunction with the \MSbar renormalization of other parameters.
Further discussions of the vev renormalization in general
theories up to the two-loop level can be found in Refs.~\cite{Sperling:2013eva,Sperling:2013xqa}.

\subsection{Lepton Sector}
The renormalization of the lepton sector of the VLL model at one-loop
order is the central part of our work. 
We begin by generating the counterterm Lagrangian with the correct structure
of renormalization constants needed to fully absorb all appearing
divergences and fulfil complete OS conditions. Afterwards, we proceed
to set up and evaluate the OS conditions on the lepton self-energies
and finally discuss the renormalization of the remaining parameters
that are defined in the $\overline{\text{MS}}$-scheme.

\subsubsection{Renormalization Transformation and Counterterm Lagrangian}
A complete counterterm Lagrangian reflecting the symmetries of the
theory is most easily obtained by a renormalization transformation of
the fields and parameters in the classical Lagrangian. To determine
the appropriate renormalization transformations, one must first choose
a specific basis for the fields and independent parameters  
like e.g.\ the gauge-eigenstate basis Eq.~\eqref{eq:L-sym} (for
the SM this was worked out in Ref.\ \cite{Bohm:1986rj}) or the
mass-eigenstate basis Eq.~\eqref{eq:L-SSB} (for the SM this was worked
out in Refs.\
\cite{Aoki:1982ed,Denner-Böhm:2001}). 
As mentioned in the Introduction, in many BSM theories, setting up an
OS renormalization scheme is significantly less straightforward than
in the SM. In our case, the main complication is the $5\times5$ lepton
mixing. There is enough freedom to treat all masses as free parameters
but the remaining parameters cannot be written as simple mixing
angles (in contrast to e.g.\ the 2HDM) and are affected by the
redundancy Eq.~\eqref{eq:wlog-transformation} between fields and parameters.

In our setup we 
follow the approach outlined in Secs.\ 4.2 and
4.3 of Ref. \cite{Hollik:2002mv} and used for the renormalization of
the chargino and neutralino sector of the MSSM in
Ref.\ \cite{Fritzsche:2002bi}. 
This approach is advantageous because it does not require a
renormalization of the mixing matrices and allows for a straight
forward combination of OS and \MSbar renormalization conditions. 

The approach proceeds in two steps. First, the
fields and parameters in the gauge-eigenstate basis are
renormalized. Secondly, an additional finite transformation
of only the fields into mass-eigenstate fields is performed. In
combination, a redundancy-free counterterm Lagrangian is obtained which
contains renormalization constants of fundamental parameters and
matrix-valued field renormalization constants for mass-eigenstate
fields, that allow to fulfil complete OS conditions.
In detail, starting from the original gauge-eigenstate Lagrangian in
Eq.~\eqref{eq:L-sym} we first introduce the renormalization
transformation for the field multiplets as 
\begin{subequations}\label{eq:symmfieldRT}
	\begin{alignat}{2}
		l^{a}_L &\to (Z^l_L)^\frac{1}{2}_{ab} \,l^b_L, \qquad &&L_R \to (Z_R^L)^\frac{1}{2} L_R,\\
		e^{a}_R &\to (Z^e_R)^\frac{1}{2}_{ab} \,e^b_R, \qquad &&E_L \to (Z_L^E)^\frac{1}{2} E_L,
	\end{alignat}
\end{subequations}
where matrix-valued field renormalization constants allow mixing
between fields with identical quantum numbers. The Dirac mass and
Yukawa parameters in Eq.\ \eqref{eq:L-sym} are renormalized as
\begin{subequations}
	\begin{alignat}{2}
		&M_L^a \to M_L^a + \delta M_L^a, \qquad
		&&M_E^a \to M_E^a + \delta M_E^a, \\
	  &Y_{ab} \,\to 
          \,Y_{ab} + \delta Y_{ab}, \qquad
	  &&~\bar\lambda ~~\to 
          \bar\lambda + \delta\bar\lambda.
	\end{alignat}
	\label{eq:parameter-CTs}
\end{subequations} 
At this point, as a consequence of the redundancy Eq.~\eqref{eq:wlog-transformation},
there is an ambiguity between the field and parameter counterterms.
Despite assuming the tree-level diagonalization Eq.~\eqref{eq:parametrization-symm-pars},
the same symmetry still implies an invariance of the bare Lagrangian under the following transformation of the renormalization constants
\begin{subequations}\label{eq:ct-transformation}
	\begin{align}
		(Z_L^l)^\frac{1}{2} &\to (\mathbb{1} + \delta V_L) (Z_L^l)^\frac{1}{2}, \\
		(Z_R^e)^\frac{1}{2} &\to (\mathbb{1} + \delta V_R) (Z_R^e)^\frac{1}{2}, \\
		M_L + \delta M_L &\to (\mathbb{1} + \delta V_L) (M_L + \delta M_L), \\
		M^T_E + \delta M^T_E &\to (M^T_E + \delta M^T_E) (\mathbb{1}+\delta V_R^\dagger), \\
		Y + \delta Y &\to (\mathbb{1} + \delta V_L) (Y + \delta Y)(\mathbb{1}+\delta V_R^\dagger),
	\end{align}
\end{subequations}
where $\mathbb{1} + \delta V_{L/R}$ can be arbitrary unitary matrices.
We make use of this transformation to bring the counterterm matrices
into the same form as the tree-level couplings
in Eq.~\eqref{eq:parametrization-symm-pars}, i.e.\ we assume that the 13
independent parameter renormalization constants in the sector have the
form 
\begin{align}\label{eq:wlog-ct}
	\delta M_L^a = (0,0,0,\delta m_L)^T, \qquad \delta M^a_E =
        (0,0,0,\delta m_E)^T, \qquad \delta Y_{ab} = \begin{pmatrix}
          \delta y_i \delta_{ij} & \delta\lambda^E_i
          \\ \delta\lambda^L_j & \delta\lambda \end{pmatrix},
        \qquad
         \delta\bar\lambda.
\end{align}
This removes the ambiguity and disentangles field and parameter counterterms. 
However, it is important to keep this choice in mind when comparing to results in the literature. 
For example, for the calculation of the $\beta$-functions of general gauge theories in Refs. \cite{Luo:2002ti,Schienbein:2018fsw} 
this ambiguity was removed by assuming hermitian field renormalization matrices, which comes at the cost of off-diagonal counterterms
$\delta M_{L/E}^i$ and $\delta y_{ij}$. We will discuss the consequences and our treatment in more detail in Sec.\ \ref{sec:RGE}.

This structure of the renormalization constants corresponding to
the gauge-eigenstate basis is sufficient to cancel all UV divergences
but not sufficient to fulfil
full OS conditions. Hence, as a second step
we perform a finite transformation to mass-eigenstate fields. We
require that the diagonalization condition in Eq.~\eqref{eq:lepton-mass-diagonalization} holds exactly, 
i.e.\ at all orders for the renormalized mass matrix. This provides an
exact 
definition of the unitary matrices $U_{L,R}$, $U^{\nu}_{L,R}$ as well
as of the mass eigenvalues $m_a$ as 
functions of the fundamental renormalized parameters. In contrast,
Eq.~\eqref{eq:mass-basis} cannot be required at all orders if full OS
conditions are desired. Instead, the
mass-eigenstate fields are defined via the following higher-order
generalization of Eq.~\eqref{eq:mass-basis},
\begin{equation}\label{eq:mefieldDef}
	\begin{pmatrix} e^a_L \\ E_L \end{pmatrix} = U_L\mathcal{R}_L\hat e_L, \quad 
	\begin{pmatrix} e^a_R \\ L^-_R \end{pmatrix} = U_R\mathcal{R}_R\hat e_R, \quad
	\begin{pmatrix}\nu^a_L \\ 0 \end{pmatrix} = U_L^{\nu}\mathcal{R}_L^{\nu}\hat\nu_L, \quad 
	\begin{pmatrix}	0 \\ L^0_R \end{pmatrix} = U_R^{\nu}\mathcal{R}_R^{\nu}\hat\nu_R.
\end{equation}
Here, we have introduced additional finite and invertible matrices
$\mathcal{R}^{(\nu)}_{L/R}=\mathbb{1} + \delta
\mathcal{R}^{(\nu)}_{L/R}$
(see also Sec. 4.2 of Ref. \cite{Hollik:2002mv}).
In total, the combination of the field transformations
(\ref{eq:symmfieldRT}) and (\ref{eq:mefieldDef}) amounts to a
renormalization transformation from bare gauge-eigenstate fields to
renormalized mass-eigenstate fields in terms of products of three matrices. Like in
Refs.\ \cite{Hollik:2002mv,Fritzsche:2002bi} it is useful rewrite
these products in terms of final field
renormalization constants which effectively renormalize
mass-eigenstate fields. Thus, the ultimate form of the
renormalization transformation from gauge-eigenstate fields to
mass-eigenstate fields is written as
\begin{subequations}\label{eq:ren-transf-mass-basis-fields}
	\begin{alignat}{2}
		\begin{pmatrix} e^a_L \\ E_L \end{pmatrix} &\to \begin{pmatrix} (Z^l_L)^\frac{1}{2} & 0 \\ 0 & (Z_L^E)^\frac{1}{2} \end{pmatrix} U_L \mathcal{R}_L \hat e_L &&\equiv U_L(Z_L)^\frac{1}{2} \hat e_L, \\
		\begin{pmatrix}	e^a_R \\ L^-_R \end{pmatrix} &\to \begin{pmatrix} (Z^e_R)^\frac{1}{2} & 0 \\ 0 & (Z_R^L)^\frac{1}{2} \end{pmatrix} U_R \mathcal{R}_R \hat e_R &&\equiv U_R (Z_R)^\frac{1}{2} \hat e_R, \\
		\begin{pmatrix}\nu^a_L \\ 0 \end{pmatrix} &\to \hspace{.44cm}\begin{pmatrix} (Z^l_L)^\frac{1}{2} & 0~ \\ 0 & 1~ \end{pmatrix} \hspace{.44cm}U_L^\nu \mathcal{R}_L^\nu \hat\nu_L && \equiv U_L^\nu(Z_L^\nu)^\frac{1}{2} \hat \nu_L, \\
		\begin{pmatrix}	0 \\ L^0_R \end{pmatrix} &\to \hspace{.35cm}\begin{pmatrix} ~\mathbb{1} & 0 \\ ~0 & (Z_R^L)^\frac{1}{2} \end{pmatrix}\hspace{.3cm} U_R^\nu \mathcal{R}_R^\nu \hat\nu_R && \equiv U_R^\nu(Z_R^\nu)^\frac{1}{2} \hat \nu_R,
	\end{alignat}
\end{subequations}
where we have introduced the four $5\times 5$ matrices $Z_{L/R}$, $Z_{L/R}^{\nu}$ which effectively implement the
matrix-valued field renormalization of the mass-eigenstate fields. We reiterate that the diagonalization
matrices $U_{L/R}$, $U_{L/R}^{\nu}$ split off on the left always remain of tree-level order. \newline

This completes the setup of the renormalization transformation. In
summary, the transformations are given by Eq.~\eqref{eq:parameter-CTs}
for the fundamental parameters (with the constraints
\eqref{eq:wlog-transformation}, \eqref{eq:wlog-ct}) and by 
Eq.~\eqref{eq:ren-transf-mass-basis-fields} (far right) for the
fields.
This renormalization transformation is applied to the
gauge-eigenstate basis 
Lagrangian Eq.~\eqref{eq:L-sym}, using the definition of the
diagonalization matrices  $U_{L/R}^{(\nu)}$ in Eq.~\eqref{eq:lepton-mass-diagonalization}.
The result is the correct structure of the bare Lagrangian expressed
in terms of renormalized fundamental parameters and mass-eigenstate
fields. It can be written as
\begin{align}\label{eq:L-ct}
	\begin{split}
		\L &\supset \bar{\hat e}^a \bigg\{i\slashed{\partial} \Big[ Z_L^{\frac{1}{2}\dagger} Z_L^\frac{1}{2} \PL + Z_R^{\frac{1}{2}\dagger} Z_R^\frac{1}{2} \PR\Big]_{ab}
		- \Big[ Z_R^{\frac{1}{2}\dagger}(m+\delta m)^\dagger Z_L^\frac{1}{2}\PL + Z_L^{\frac{1}{2}\dagger}(m+\delta m) Z_R^\frac{1}{2}\PR \Big]_{ab} \\
		&\hspace{.55cm} - \tfrac{1}{\sqrt{2}}\Big[Z_R^{\frac{1}{2}\dagger}(Y^h +\delta Y^h)^\dagger Z_L^\frac{1}{2} \PL + Z_L^{\frac{1}{2}\dagger}(Y^h+\delta Y^h) Z^\frac{1}{2}_R\Big]_{ab} Z_h h \bigg\} \hat e^b \\
		& + \bar{\hat{\nu}}^a \bigg\{ i\slashed{\partial} \Big[ Z_L^{\nu\frac{1}{2}\dagger} Z_L^{\nu\frac{1}{2}} \PL + Z_R^{\nu\frac{1}{2}\dagger} Z_R^{\nu\frac{1}{2}} \PR\Big]_{ab} 
		-  (m_L+\delta m_L)\Big[Z^{\nu\frac{1}{2}\dagger}_{R,a4} Z^{\nu\frac{1}{2}}_{L,4b} \PL + Z^{\nu\frac{1}{2}\dagger}_{L,a4} Z^{\nu\frac{1}{2}}_{R,4b} \PR \Big] \bigg\}\hat\nu^b,
	\end{split}
\end{align}
which includes a bare counterpart of the tree-level Lagrangian Eq.~\eqref{eq:L-SSB},
where the mass and Yukawa matrices effectively renormalize as 
\begin{subequations}\label{eq:mass-basis-par-cts}
	\begin{alignat}{2}
		m_a \delta_{ab} + \delta m_{ab} &= m_a \delta_{ab} +  \left[ U_L^\dagger ~\begin{pmatrix} \delta (Yv) & \delta M_L \\ \delta M^T_E & \delta (\lambar v) \end{pmatrix} U_R \right]_{ab} \\
		Y^h_{ab} + \delta Y^h_{ab} ~~&= ~\,Y^h_{ab} ~~+ \left[ U_L^\dagger  ~\begin{pmatrix} \delta Y & 0~ \\ 0 & \delta \lambar ~\end{pmatrix} U_R \right]_{ab} &&
	\end{alignat}
\end{subequations}
We reiterate that the fundamental parameters are taken as the ones of Eq.~\eqref{eq:L-sym} and the fundamental renormalization constants are the ones of Eq.~\eqref{eq:parameter-CTs} and 
Eq.~\eqref{eq:ren-transf-mass-basis-fields} (far right). Other quantities, in particular $U_{L/R}$ and $m_a$ as well as the renormalization constants 
$\delta m_{ab}$ and $\delta Y^h$ are derived from these definitions and are given by Eq.~\eqref{eq:lepton-mass-diagonalization}
and Eq.~\eqref{eq:mass-basis-par-cts}, respectively.

\subsubsection{One-loop On-Shell Renormalization}
Having established the structure of the counterterm Lagrangian we set out to solve the OS renormalization conditions on the lepton self-energies 
for the renormalization constants at one-loop order. In terms of the
renormalized charged lepton self-energy matrix $\hat\Sigma_{ab}$ these conditions read
\begin{equation}\label{eq:lep-OS-condition-1}
	\re \hat\Sigma_{ab}(\slashed{p}) u_b(p) \big|_{p^2 = m_b^2} = 0, \qquad \quad 
	\lim_{p^2\to m_a^2} \tfrac{1}{\slashed{p}-m_a} \re \hat\Sigma_{aa}(\slashed{p}) u_a(p) = 0, \qquad a,b=1,...,5
\end{equation}
where $\re$ discards the imaginary parts resulting from the loop integration.
To solve for the renormalization constants we first write the renormalized self-energy in terms of the  covariant decomposition
\begin{align}\label{eq:renorm-SE}
	i\hat\Sigma_{ab}(\slashed{p}) = i\slashed{p}\Big(\hat\Sigma^R_{ab}(p^2)\PR + \hat\Sigma_{ab}^L(p^2)\PL \Big) + i\Big(\hat\Sigma_{ab}^{SR}(p^2)\PR + \hat\Sigma_{ab}^{SL}(p^2)\PL\Big),
\end{align}
where hermiticity of the (counterterm) Lagrangian relates the coefficient functions by
\begin{align}
	\re\hat\Sigma^{R}_{ab}(p^2) = \re\hat\Sigma^{R*}_{ba}(p^2), \quad \re\hat\Sigma^{L}_{ab}(p^2) = \re\hat\Sigma^{L*}_{ba}(p^2), \quad \re\hat\Sigma^{SR}_{ab}(p^2) = \re\hat\Sigma^{SL*}_{ba}(p^2).
\end{align}
Next, we expand the charged lepton field renormalization matrices in Eq.~\eqref{eq:L-ct} to one-loop order
\begin{align}
	Z^{\frac{1}{2}}_{L/R} \approx \mathbb{1} + \tfrac{1}{2} \delta Z_{L/R}.
\end{align}
The covariant coefficients in Eq.~\eqref{eq:renorm-SE} can now be expressed in terms of the corresponding coefficients of the unrenormalized one-loop self-energy matrix $\Sigma_{ab}$ 
and the one-loop counterterm contributions
\begin{subequations}
	\begin{align}
		\hat\Sigma^R_{ab}(p^2) &= \Sigma^R_{ab}(p^2) + \tfrac{1}{2}(\delta Z_R)_{ab} + \tfrac{1}{2}(\delta Z_R^\dagger)_{ab} \\
		\hat\Sigma^L_{ab}(p^2) &= \Sigma^L_{ab}(p^2) + \tfrac{1}{2}(\delta Z_L)_{ab} + \tfrac{1}{2}(\delta Z_L^\dagger)_{ab} \\
		\hat\Sigma^{SR}_{ab}(p^2) &= \Sigma^{SR}_{ab}(p^2) - \tfrac{1}{2} m_a(\delta Z_R)_{ab} - \tfrac{1}{2}(\delta Z_L^\dagger)_{ab} m_b - (\delta m)_{ab}\\
		\hat\Sigma^{SL}_{ab}(p^2) &= \Sigma^{SL}_{ab}(p^2) - \tfrac{1}{2} m_a(\delta Z_L)_{ab} - \tfrac{1}{2}(\delta Z_R^\dagger)_{ab} m_b - (\delta m^\dagger)_{ab}.
	\end{align}
\end{subequations}
Finally, the OS conditions Eq.~\eqref{eq:lep-OS-condition-1} can be solved and imply the following expression for the renormalization constants of the charged leptons
\begin{subequations}\label{eq:OS-charged-lepton-rc}
	\begin{align}
		&\text{for~} a=b \quad
		\begin{cases}
				(\delta m)_{aa} &= \frac{1}{2}\,\re \bigg[m_a \Sigma^R_{aa} + m_a \Sigma^L_{aa} + \Sigma^{SR}_{aa} + \Sigma^{SL}_{aa}\bigg]\Big|_{p^2=m_a^2} \\[.3cm]
				(\delta Z_R)_{aa} &= -\re\bigg[\Sigma^R_{aa} + m_a\Big(\Sigma^{SR\prime}_{aa} + \Sigma^{SL\prime}_{aa}\Big)
				+ m_a^2\Big(\Sigma^{R\prime}_{aa} + \Sigma^{L\prime}_{aa}\Big) \bigg]\Big|_{p^2=m_a^2} \\[.3cm]
				(\delta Z_L)_{aa} &= -\re\bigg[\Sigma^L_{aa} + m_a\Big(\Sigma^{SR\prime}_{aa} + \Sigma^{SL\prime}_{aa}\Big)
				+ m_a^2\Big(\Sigma^{R\prime}_{aa} + \Sigma^{L\prime}_{aa}\Big)\bigg]\Big|_{p^2=m_a^2}
			\end{cases}\label{eq:cL-diagonal-cts}	 \\[.3cm]
		&\text{for~} a\neq b \quad
		\begin{cases}
				(\delta Z_R)_{ab} &= \frac{2}{m_a^2 - m_b^2} \bigg[m_b^2 \re\Sigma^R_{ab} + m_a m_b \re\Sigma^L_{ab}
				+ m_a \re\Sigma^{SR}_{ab} \\
				&\hspace{2.1cm} + m_b\re\Sigma^{SL}_{ab} - m_a(\delta m)_{ab} - m_b (\delta m^\dagger)_{ab} \bigg]\Big|_{p^2=m_b^2} \\[.3cm]
				(\delta Z_L)_{ab} &= \frac{2}{m_a^2 - m_b^2} \bigg[m_b^2 \re\Sigma^L_{ab} + m_a m_b \re\Sigma^R_{ab}
				+ m_a\re\Sigma^{SL}_{ab} \\
				&\hspace{2.1cm} + m_b\re\Sigma^{SR}_{ab} - m_a(\delta m^\dagger)_{ab} - m_b(\delta m)_{ab} \bigg]\Big|_{p^2=m_b^2}
			\end{cases}
	\end{align}
\end{subequations}
The neutral lepton sector can be treated analogously. The main difference is
that the structure of the theory guarantees that there are three
massless and one massive neutrino and the mixing is trivial. Hence, the
The corresponding conditions on the neutrino self-energies result in
\begin{subequations}\label{eq:OS-neutral-lepton-rc}
	\begin{alignat}{2}
			&\text{for~} a=b
			&&\begin{cases}
					\delta m_L &= \frac{1}{2}\,\re \bigg[m_L \Sigma^{\nu R}_{44} + m_L \Sigma^{\nu L}_{44} + \Sigma^{\nu SR}_{44} + \Sigma^{\nu SL}_{44}\bigg]\Big|_{p^2=m_L^2} \\[.3cm]
					(\delta Z^\nu_R)_{aa} &= -\re\bigg[\Sigma^{\nu R}_{aa} + \delta_{a4} m_L \Big(\Sigma^{\nu SR\prime}_{aa} + \Sigma^{\nu SL\prime}_{aa} + 
					m_L\Sigma^{\nu R\prime}_{aa} + m_L\Sigma^{\nu L\prime}_{aa}\Big) \bigg]\Big|_{p^2=\delta_{a4}m_L^2} \\[.3cm]
					(\delta Z^\nu_L)_{aa} &= -\re\bigg[\Sigma^{\nu L}_{aa} + \delta_{a4} m_L \Big(\Sigma^{\nu SR\prime}_{aa} + \Sigma^{\nu SL\prime}_{aa}
					+ m_L \Sigma^{\nu R\prime}_{aa} + m_L\Sigma^{\nu L\prime}_{aa}\Big) \bigg]\Big|_{p^2=\delta_{a4}m_L^2}
				\end{cases}\label{eq:nL-diagonal-cts}	 \\[.3cm]
			&\text{for~} a,b \neq 4
			&&\begin{cases}
					(\delta Z^\nu_{L/R})_{a4} &= - \frac{2}{m_L} \re\bigg[m_L\Sigma^{\nu L/R}_{a4}(m_L^2) + \Sigma^{\nu SR/SL}_{a4}(m_L^2)\bigg] \\[.3cm]
					(\delta Z^\nu_{L/R})_{4b} &= \hphantom{-}\frac{2}{m_L} \re \Sigma^{\nu SL/SR}_{4b}(0)
			\end{cases}.
		\end{alignat}
\end{subequations}
Note that the OS conditions on the off-diagonal neutrino self-energies
for $a,b\neq 4$ are fulfilled automatically and do not provide any
non-trivial constraints for renormalization constants. 

\subsubsection{Renormalization of the Fundamental Parameters}
The OS conditions listed above do not fix all fundamental renormalization constants of Eq.~\eqref{eq:parameter-CTs} and 
Eq.~\eqref{eq:ren-transf-mass-basis-fields} unambiguously.
The remaining freedom has to be fixed by imposing additional conditions. In order to motivate our choice we start by writing the relation Eq.~\eqref{eq:mass-basis-par-cts}
for the diagonal mass renormalization constants as
\begin{align}\label{eq:OS-ct-relation}
	\begin{split}
		\delta m_{aa} &= \sum_{b, c} (U_L^\dagger)_{ab} (U_R)_{ca} \delta M_{bc}, \qquad a = 1,...,5, \qquad
		\delta M = \begin{pmatrix} \delta (Yv) & \delta M_L \\ \delta M_E^{\text{T}} & \delta (\lambar v) \end{pmatrix}.
	\end{split}
\end{align} 
The OS conditions fix the left-hand side for all $a=1\ldots5$
via Eq.~\eqref{eq:cL-diagonal-cts}.
In addition, the neutral lepton OS condition \eqref{eq:nL-diagonal-cts}
directly fixes $\delta m_L$. 
Hence,  6  of the 13 renormalization constants $\delta m_L$, $\delta
m_E$, $\delta Y_{ab}$ and $\delta \lambar$ can be determined from
OS conditions
but further conditions need to be imposed on the remaining
renormalization constants.
Some care has to be taken in order to avoid absorbing large corrections of order of the VLL masses $m_{4,5} \gg v$ into e.g.\ $\delta (\lambda v)$ or $\delta (\lambar v)$.
Such corrections are instead naturally absorbed by $\delta m_E$ and $\delta m_L$. In principle, this implies we should solve Eq.~\eqref{eq:OS-ct-relation}
for $\delta m_L$ and $\delta m_E$ in addition to 3 of the Yukawa counterterms. However, since $\delta m_L$ is already fixed by the neutrino OS conditions we are forced to choose
a fourth Yukawa counterterm instead. Fortunately, the difference between $\delta m_L$ obtained from the neutral lepton OS conditions 
and from solving Eq.~\eqref{eq:OS-ct-relation}, like the difference between $m_4$ and $m_L$, is generated by EWSB effects. It is therefore of order $v$ and can
safely be absorbed into e.g.\ $\delta (\lambda v)$ or $\delta (\lambar v)$. An obvious choice for the four Yukawa counterterms is $\delta y_i$ (to allow for the OS renormalized SM limit) 
and either $\delta \lambda$ or $\delta\lambar$. Here we choose $\delta \lambda$, such that Eq.~\eqref{eq:OS-ct-relation} has to be solved for $\delta (Y_{aa}v)$ and $\delta m_E$.
For simplicity, we fix the remaining renormalization constants in the \MSbar scheme. 
To summarize, the parameter counterterms are determined in the following way:
\begin{itemize}
	\item $\delta m_E$, $\delta (y_i v)$ and $\delta (\lambda v)$
          are fixed via the charged lepton masses by Eq.~\eqref{eq:OS-ct-relation} together with Eq.~\eqref{eq:cL-diagonal-cts}
	\item $\delta m_L$ is fixed via the neutrino mass by Eq.~\eqref{eq:nL-diagonal-cts}
	\item $\delta \lambda^L_i,  \delta \lambda^E_i$ and $\delta \lambar$ are fixed by the \MSbar scheme
	\item $\delta v$ is fixed by Eq.~\eqref{eq:dv} and the tadpole
          is fixed by Eq.~\eqref{eq:tadpoleDef}.
\end{itemize}
This provides sufficient conditions to fix all $\delta M_{ab}$ and in turn all $\delta m_{ab}$ and $\delta Y^h_{ab}$ unambiguously.
To solve for the set of renormalization constants, we first write Eq.~\eqref{eq:OS-ct-relation} in matrix notation
\begin{align}
	\bm{\delta} \mathbf{m} = \bm{\kappa} \cdot \bm{\delta} \mathbf{M} + \bm{\delta}\mathbf{u}, \qquad \text{where} \quad \bm{\delta}\mathbf{M} = \{\delta (y_1 v),\delta (y_2 v),\delta (y_3 v), \delta (\lambda v), \delta m_E\}, 
\end{align}
with the coefficient matrix $\bm\kappa$ and constant vector $\bm{\delta}\mathbf{u}$ given by 
\begin{align}\label{eq:kappa-definition}
	\bm{\kappa}_{ab} =\begin{cases}
			(U_L^\dagger)_{ab} (U_R)_{ba} & b = 1,...,4\\ 
			(U_L^\dagger)_{a5} (U_R)_{4a} & b=5
		\end{cases}, \qquad \bm{\delta}\mathbf{u}_a = (U_L^\dagger)_{a5} (U_R)_{5a} \delta M_{55} + \sum_{\substack{b\neq c \\ b \neq 5}} (U_L^\dagger)_{ab} (U_R)_{ca} \delta M_{bc}.
\end{align}
This allows us to write the solution in the following compact form
\begin{align}\label{eq:inverted-cts}
	\bm{\delta} \mathbf{M} = \bm{\kappa}^{-1} (\bm {\delta} \mathbf{m} - \bm{\delta}\mathbf{u} ), \qquad \delta Y_{aa} = \tfrac{1}{v} \big(\delta M_{aa} - Y_{aa} \delta v\big).
\end{align}
These equations hold separately for the finite and UV divergent parts
and can be evaluated numerically or using perturbative expressions for the rotation matrices such as those given in \cite{Dermisek:2013gta}.
The remaining \MSbar parameter counterterms are taken from the literature \cite{Luo:2002ti,Schienbein:2018fsw}.

\subsection{Relation to \MSbar and Renormalization Group Running}\label{sec:RGE}
In the renormalization scheme defined by the renormalization
transformations Eq.~\eqref{eq:parameter-CTs} and
Eq.~\eqref{eq:ren-transf-mass-basis-fields} (right),
the on-shell conditions Eq.~\eqref{eq:lep-OS-condition-1} and by the relation Eq.~\eqref{eq:inverted-cts},
the mass-eigenstate fields are on-shell renormalized and the diagonalized mass matrices Eq.~\eqref{eq:lepton-mass-diagonalization} correspond to their respective physical
pole masses.
However, in this way only a subset of the fundamental parameters are determined by physical on-shell conditions while the rest are defined in
the \MSbar scheme with the constraint
\eqref{eq:wlog-ct}. Here we describe how to determine the
values of the 
\MSbar counterterms and the resulting running of all renormalized
parameters.
We begin with the well-known $\beta$-functions derived for general gauge theories 
in the \MSbar scheme, see e.g.\ \cite{Schienbein:2018fsw} and references therein. As mentioned in the previous section, these $\beta$-functions are derived
assuming hermitian field renormalization matrices which resolves the redundancy between field and parameter renormalization. We instead removed this redundancy by imposing
Eq.~\eqref{eq:wlog-ct} on the parameter counterterms, allowing for general field renormalization matrices. As a result, the renormalization constants including the $\frac{1}{\epsilon}$ poles,
and consequently the $\beta$-functions, change. We have explicitly computed the required matrices $\delta V_{L/R}$ of Eq.~\eqref{eq:ct-transformation} that absorb
the off-diagonal parameter counterterms into the field renormalization matrices. The resulting $\beta$-functions corresponding to \MSbar with condition Eq.~\eqref{eq:wlog-ct} are listed
in App.~\ref{App:RGE}.
These $\beta$-functions unambiguously determine the \MSbar
renormalization constants $\delta\lambda^E_i$, $\delta\lambda^L_j$,
$\delta\lambar$ required in our scheme.

The running of parameters in our (partial) OS scheme is further modified by the presence of finite contributions 
$\delta\Theta_\alpha^{[0]}$ to the renormalization constants $\delta\Theta_\alpha=\frac{1}{\epsilon}\delta\Theta^{[1]}_\alpha+\delta\Theta^{[0]}_\alpha$ 
for some of the fundamental model parameters collectively denoted by $\Theta_\alpha$. Specifically, for our choices in the previous section,  
these finite contributions relate the model parameters in our scheme to the ones in the pure \MSbar scheme by
\begin{align}\label{eq:OS-to-MS}
	m_L^{\overline{\text{MS}}} = m_L + \delta m^{[0]}_L, \qquad m_E^{\overline{\text{MS}}} = m_E + \delta m^{[0]}_E, \qquad Y_{aa}^{\overline{\text{MS}}} = Y_{aa} + \delta Y_{aa}^{[0]}
\end{align}
The resulting running of the parameters $\Theta_\alpha$ in our scheme is therefore given by
\begin{align}\label{eq:OS-RGE}
	\ddmu{\Theta_\alpha}  \equiv \beta^\text{eff}(\Theta _\alpha) \approx \beta(\Theta_\alpha) - \ppmu{\delta\Theta_\alpha^{[0]}} + \O(2L)
\end{align}
For purely OS renormalized parameters, like $m_L$, the effective $\beta$-function vanishes. However, because of the mixing between OS and \MSbar parameters
when going to the mass basis the parameters, $Y_{aa}$ and $m_E$ necessarily retain some residual $\mu$ dependence. 
We also note that due to the mixing between \MSbar and OS counterterms in Eq.~\eqref{eq:inverted-cts} the usual (one-loop) OS relation
\begin{align}
	\mu\frac{\partial \delta\Theta_\alpha^{[0]}}{\partial \mu} = 2 \delta \Theta_\alpha^{[1]}
\end{align}
does not hold for these parameters.

\section{\boldmath One-Loop Effective Higgs Coupling and Correlation with $a_\mu$}\label{sec:one-loop}

Having established a renormalization scheme that addresses the
higher-order mixing between the lepton fields in the VLL model and
establishes OS conditions on masses and fields, in this section we 
test its validity and discuss a first application to the
phenomenological impact of higher-order corrections.

In particular, we focus on the effective muon--Higgs coupling
$\lambda_{\mu\mu}$ in the muon-specific case,
where the VLL couplings to the first and third generation of SM
leptons are zero, $\lambda^E_{1,3}=\lambda^L_{1,3}=0$.
The effective coupling $\lambda_{\mu\mu}$ is one of the most
interesting observables in the VLL model as it receives large 
chirally enhanced tree-level corrections from the mixing between SM and VL leptons. Furthermore, these corrections are directly correlated with the one-loop
VLL contributions to the anomalous magnetic moment $a_\mu$. For this
reason the tree-level effective coupling has been studied extensively
in the  
muon-specific VLL model in search for an explanation of the tentative experimental deviation of $\Delta a_\mu$ from the SM prediction \cite{Kannike:2011ng,Dermisek:2013gta,Dermisek:2020cod,Dermisek:2023tgq}.
Here, we calculate the one-loop corrections to $\lambda_{\mu\mu}$, and
we show how these higher-order contributions can significantly modify
the correlation between $\lambda_{\mu\mu}$ and $\Delta a_\mu^\text{VLL}$ found at
tree-level, see Eq.~\eqref{eq:amuVLL}.

\subsection{Definition and Vertex Renormalization}
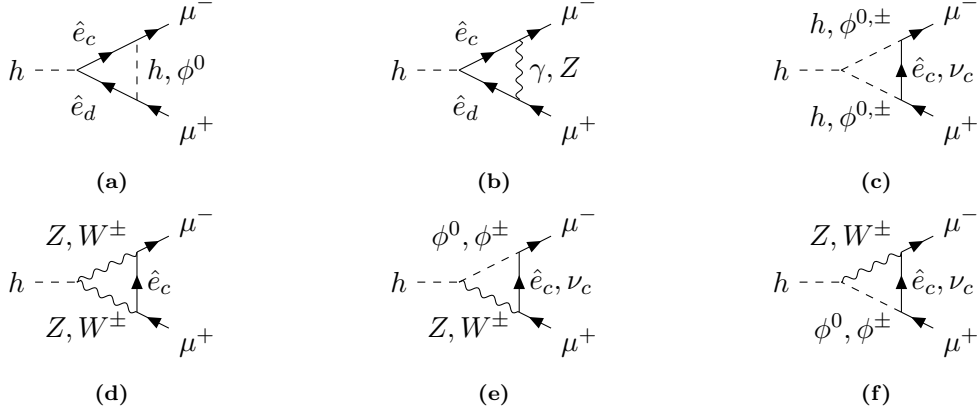
\begin{figure}
	\centering
	\begin{subfigure}[b]{.3\textwidth}
		\centering
		\begin{tikzpicture}[baseline=-.5ex, scale=.8]
			\begin{feynman}[small]
				\vertex (h) at (-1.5,0) {$h$};
				\vertex (m1) at (1.5,1) {$\mu^-$};
				\vertex (m2) at (1.5,-1) {$\mu^+$};
				\vertex (a) at (-.5,0);
				\vertex (b) at (.5,.5);
				\vertex (c) at (.5,-.5);
				\diagram*{
					(h) -- [scalar] (a) -- [fermion, edge label={$\hat e_c$}] (b) -- [fermion] (m1),
					(m2) -- [fermion] (c) -- [fermion, edge label={$\hat e_d$}] (a),
					(c) -- [scalar, edge label'={$h, \phi^0$}] (b)
				};
			\end{feynman}
		\end{tikzpicture}
		\caption{}
	\end{subfigure}
	\begin{subfigure}[b]{.3\textwidth}
		\centering
		\begin{tikzpicture}[baseline=-.5ex, scale=.8]
			\begin{feynman}[small]
				\vertex (h) at (-1.5,0) {$h$};
				\vertex (m1) at (1.5,1) {$\mu^-$};
				\vertex (m2) at (1.5,-1) {$\mu^+$};
				\vertex (a) at (-.5,0);
				\vertex (b) at (.5,.5);
				\vertex (c) at (.5,-.5);
				\diagram*{
					(h) -- [scalar] (a) -- [fermion, edge label={$\hat e_c$}] (b) -- [fermion] (m1),
					(m2) -- [fermion] (c) -- [fermion, edge label={$\hat e_d$}] (a),
					(c) -- [boson, edge label'={$\gamma, Z$}] (b)
				};
			\end{feynman}
		\end{tikzpicture}		
		\caption{}
	\end{subfigure}
	\begin{subfigure}[b]{.3\textwidth}
		\centering
		\begin{tikzpicture}[baseline=-.5ex, scale=.8]
			\begin{feynman}[small]
				\vertex (h) at (-1.5,0) {$h$};
				\vertex (m1) at (1.5,1) {$\mu^-$};
				\vertex (m2) at (1.5,-1) {$\mu^+$};
				\vertex (a) at (-.5,0);
				\vertex (b) at (.5,.5);
				\vertex (c) at (.5,-.5);
				\diagram*{
					(h) -- [scalar] (a) -- [scalar, edge label={$h,\phi^{0,\pm}\hspace{-.5cm}$},inner sep=6pt] (b) -- [fermion] (m1),
					(m2) -- [fermion] (c) -- [scalar, edge label={$h,\phi^{0,\pm}\hspace{-.5cm}$}, inner sep=6pt] (a),
					(c) -- [fermion, edge label'={$\hat e_c,\nu_c$}] (b)
				};
			\end{feynman}
		\end{tikzpicture}
		\caption{}
	\end{subfigure}
	\begin{subfigure}[b]{.3\textwidth}
		\centering
		\begin{tikzpicture}[baseline=-.5ex, scale=.8]
			\begin{feynman}[small]
				\vertex (h) at (-1.5,0) {$h$};
				\vertex (m1) at (1.5,1) {$\mu^-$};
				\vertex (m2) at (1.5,-1) {$\mu^+$};
				\vertex (a) at (-.5,0);
				\vertex (b) at (.5,.5);
				\vertex (c) at (.5,-.5);
				\diagram*{
					(h) -- [scalar] (a) -- [boson, edge label={$Z,W^\pm\hspace{-.5cm}$},inner sep=6pt] (b) -- [fermion] (m1),
					(m2) -- [fermion] (c) -- [boson, edge label={$Z,W^\pm\hspace{-.5cm}$},inner sep=6pt] (a),
					(c) -- [fermion, edge label'={$\hat e_c$}] (b)
				};
			\end{feynman}
		\end{tikzpicture}
		\caption{}
	\end{subfigure}
	\begin{subfigure}[b]{.3\textwidth}
		\centering
		\begin{tikzpicture}[baseline=-.5ex, scale=.8]
			\begin{feynman}[small]
				\vertex (h) at (-1.5,0) {$h$};
				\vertex (m1) at (1.5,1) {$\mu^-$};
				\vertex (m2) at (1.5,-1) {$\mu^+$};
				\vertex (a) at (-.5,0);
				\vertex (b) at (.5,.5);
				\vertex (c) at (.5,-.5);
				\diagram*{
					(h) -- [scalar] (a) -- [scalar, edge label={$\phi^0,\phi^\pm\hspace{-.5cm}$},inner sep=6pt] (b) -- [fermion] (m1),
					(m2) -- [fermion] (c) -- [boson, edge label={$Z,W^\pm\hspace{-.5cm}$},inner sep=6pt] (a),
					(c) -- [fermion, edge label'={$\hat e_c,\nu_c$}] (b)
				};
			\end{feynman}
		\end{tikzpicture}
		\caption{}
	\end{subfigure}
	\begin{subfigure}[b]{.3\textwidth}
		\centering
		\begin{tikzpicture}[baseline=-.5ex, scale=.8]
			\begin{feynman}[small]
				\vertex (h) at (-1.5,0) {$h$};
				\vertex (m1) at (1.5,1) {$\mu^-$};
				\vertex (m2) at (1.5,-1) {$\mu^+$};
				\vertex (a) at (-.5,0);
				\vertex (b) at (.5,.5);
				\vertex (c) at (.5,-.5);
				\diagram*{
						(h) -- [scalar] (a) -- [boson, edge label={$Z,W^\pm\hspace{-.5cm}$},inner sep=6pt] (b) -- [fermion] (m1),
						(m2) -- [fermion] (c) -- [scalar, edge label={$\phi^0,\phi^\pm\hspace{-.5cm}$},inner sep=6pt] (a),
						(c) -- [fermion, edge label'={$\hat e_c,\nu_c$}] (b)
					};
			\end{feynman}
		\end{tikzpicture}
		\caption{}
	\end{subfigure}
	\caption{One-loop virtual contributions to $h \to \mu^+ \mu^-$ in the VLL model.}
	\label{fig:h-to-mu-mu}
\end{figure}
At tree-level, the effective muon--Higgs coupling $\lambda_{\mu\mu}$ can be directly defined by the partial decay width of an on-shell Higgs into a muon/anti-muon pair
\begin{align}
	\Gamma(h\to \mu^- \mu^+) \approx \frac{M_h}{8\pi^2} |\lambda_{\mu\mu}^\text{tree}|^2 + \O(1L).
\end{align} 
Neglecting the muon mass, it is related to the tree-level matrix element by
\begin{align}
	|\lambda^\text{tree}_{\mu\mu}|^2 = \tfrac{1}{M_h^2} \overline{|\M|^2}(h\to \mu^-\mu^+)_\text{tree} = (Y^h_{22})^2.
\end{align}
This definition can be extended to the one-loop level in the following
simple way. In general, the loop-corrected on-shell matrix element
can be written as
\begin{align}
	\M\big(h(q)\to\mu^-(p)\mu^+(p')\big) = -i \bar u(p) \Big[ \zeta(q^2,p^2,p'^2)^* \PL + \zeta(q^2,p^2,p'^2) \PR \Big] v(p').
\end{align}
At next-to-leading order (NLO) the coefficient function $\zeta$
develops an infrared (IR) divergence due to the vertex and self-energy corrections from soft photon loops. 
This divergence can be factored out into a universal gauge-invariant quantity $\alpha B$ \cite{Yennie:1961ad},
\begin{align}
	\zeta = \exp(\alpha B) \zeta_0,
\end{align}
such that $\zeta_0$ is IR finite, and where (in dimensional regularization)
\begin{align}
	\begin{split}
		\alpha B = \frac{e^2}{16\pi^2} \Big\{4 &+ B_0(0,m_\mu^2,m_\mu^2) - B_0(M_h^2,m_\mu^2,m_\mu^2) + 4 m_\mu^2 B'_0(m_\mu^2,0,m_\mu^2)\\
		& -2(M_h^2-2m_\mu^2) C_0(M_h^2,m_\mu^2,m_\mu^2,m_\mu^2,m_\mu^2,0)\Big\}.
	\end{split}
\end{align}
We can thus define an IR finite effective muon--Higgs coupling at the loop-level by\footnote{It would be straightforward to extend this definition
to include the real radiation corrections that remove the IR divergences from the physical decay width. However, like the soft singularities, 
these contributions cancel in the ratio $\lambda_{\mu\mu}/\lambda_{\mu\mu}^\text{SM}$ and have no impact on our further analysis.}
\begin{align}\label{eq:lambdamumu1L}
	\lambda_{\mu\mu} \equiv \sqrt{2}\zeta_0(M_h^2,m_\mu^2,m_\mu^2) = \sqrt{2} \exp(-\alpha B) \zeta(M_h^2,m_\mu^2,m_\mu^2).
\end{align}
At one-loop order this expression can be written explicitly as the tree-level value plus the virtual corrections $\zeta^{1l}$ from the diagrams
depicted in Fig.~\ref{fig:h-to-mu-mu} and the counterterm contribution generated by the Lagrangian Eq.~\eqref{eq:L-ct}.
\begin{align}\label{eq:eff-Higgs-coupling}
	\lambda^{\text{tree}+1l}_{\mu\mu} = Y^h_{22} (1 - \alpha B) + \sqrt{2}\zeta^{1l}(M_h^2,m_\mu^2,m_\mu^2) + \delta Y^h_{22} 
	+ \tfrac{1}{2} (\delta Z_{L}^\dagger Y^h + Y^h \delta Z_{R} + \delta Z_h Y^h)_{22}.
\end{align}
To compute the numerical results, the amplitudes for the diagrams in Fig.\ \ref{fig:h-to-mu-mu} were obtained using \texttt{FeynArts} \cite{Hahn:2000kx} and 
\texttt{FeynCalc} \cite{Mertig:1990an,Shtabovenko:2016sxi,Shtabovenko:2020gxv,Shtabovenko:2023idz} and the appearing loop function were evaluated
using \texttt{LoopTools} \cite{Hahn:1998yk}. In \texttt{LoopTools} IR divergent loop functions are automatically regulated using a finite photon mass. 
We have checked that our results are independent of the artificial photon mass within the numerical precision.

Importantly, we have also confirmed numerically that the UV divergences in Eq.~\eqref{eq:eff-Higgs-coupling} cancel. 
This cancellation is non-trivial since the renormalization constants are defined using 
a combination of self-energy diagrams and explicit
${1}/{\epsilon}$ poles for the \MSbar counterterms derived from
the modified $\beta$-functions in App.~\ref{App:RGE}, 
while the UV divergences of $\zeta^{1l}$ are obtained from the genuine
3-point loop diagrams in Fig.~\ref{fig:h-to-mu-mu}. The cancellation thus constitutes a	
significant check of the renormalization scheme.

\subsection{Cancellation of the Renormalization Scale Dependence}

\begin{figure}
	\centering
	\includegraphics[width=0.45\textwidth]{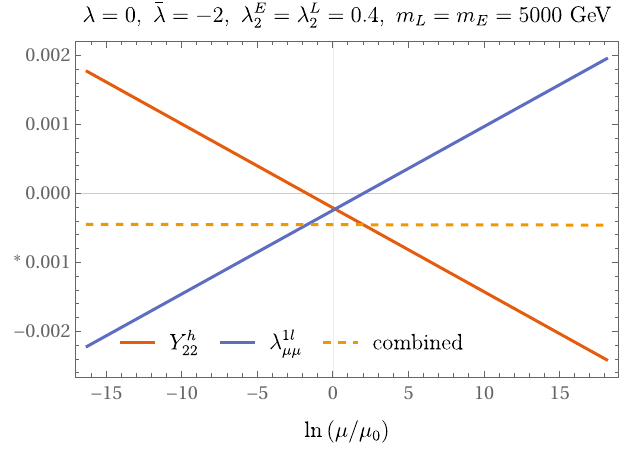}
	\caption{Example of cancellation between the explicit scale dependence of the one-loop contributions to the effective muon--Higgs coupling $\lambda_{\mu\mu}^{1l}(\mu)$ and the (linearized) running of the tree-level effective coupling $Y^h_{22}(\mu)$.}
	\label{fig:running-couplings}
\end{figure}

A second important check for the validity of the scheme is the cancellation of the renormalization scale dependence in physical observables. 
As discussed in Sec.~\ref{sec:RGE}, due to the combination of OS and
\MSbar conditions, the fundamental model parameters retain a residual 
renormalization scale $\mu$-dependence. At the same time the one-loop
contributions to the  effective Higgs coupling in
Eq.\ \eqref{eq:lambdamumu1L} depend on  the renormalization
scale $\mu$ explicitly. In total, the explicit and implicit
$\mu$-dependences must cancel exactly at the one-loop level.

The implicit $\mu$-dependence is governed by the modified renormalization group equation Eq.~\eqref{eq:OS-RGE} which can,
in principle, be solved to obtain the one-loop running of the
fundamental gauge-eigenstate parameters $Y_{ab}(\mu)$ and $m_E(\mu)$. To check the cancellation of the
scale dependence in the physical observable, i.e.\ in
Eq.\ \eqref{eq:lambdamumu1L}, we are interested in the effect of this
running on the effective coupling matrix $Y^h_{22}(\mu)$ in the mass-eigenstate basis.
By construction, the diagonalized lepton mass matrices represent the physical pole masses and are independent of $\mu$. However, the effective Yukawa
coupling matrix   inherits a scale dependence from both the running couplings and the scale dependence of the rotation matrices $U_{L/R}$. 
In general, solving the RGE and switching to the mass basis results in a complicated non-linear dependence of the coupling matrices on the 
renormalization scale.
Since the cancellation between the implicit and the explicit
$\mu$-dependence
can only be seen up to the considered loop order, it is
useful to truncate the higher-order terms analytically. Working at
one-loop order this means we should only consider the linear
dependence on $\ln(\mu)$ in both the running couplings and loop corrections. In this approximation, the renormalization scale dependence of the
fundamental parameters can be obtained directly from Eq.~\eqref{eq:OS-RGE} by integration
\begin{align}\label{eq:linearized-running}
	\Theta_\alpha(\mu) \approx \Theta_\alpha(\mu_0) +  \beta^\text{eff}(\Theta_\alpha(\mu_0)) \ln(\tfrac{\mu}{\mu_0}) + \O(2L).
\end{align}
From this, the linearized running of the effective couplings in the mass basis can be obtained analytically using the perturbative expressions
of the diagonalization matrices given in Ref.~\cite{Dermisek:2013gta}
\begin{align}
	Y^h_{22}(\mu) \approx Y^h_{22}(\mu_0) + \frac{2v^2 \lambar\lambda^E_2 \lambda^L_2}{m_4 m_5}\bigg(\frac{\beta^\text{eff}(\lambar)}{\lambar} 
	+ \frac{\beta^\text{eff}( \lambda^E_2)}{\lambda^E_2} + \frac{\beta^\text{eff}(\lambda^L_2)}{\lambda^L_2} \bigg) \ln(\tfrac{\mu}{\mu_0}) + \O(v^4).
\end{align}
Fig. \ref{fig:running-couplings} shows an example of how this linearized running of the tree-level coupling exactly cancels the explicit 
one-loop logarithms appearing in the renormalized (IR-subtracted) vertex corrections.

\subsection{Qualitative Discussion and Muon $g-2$}\label{sec:qualitative-impact}

Having confirmed the validity of the renormalization scheme, we move on to discuss the qualitative impact of the one-loop corrections.
In particular we focus on the correlation between $\lambda_{\mu\mu}$ and the VLL contributions to the muon anomalous magnetic moment $\Delta a_\mu^\text{VLL}$. 
For the calculation of $\Delta a_\mu^\text{VLL}$, our implementation
agrees with the formula presented e.g.\ in
Ref.\ \cite{Dermisek:2013gta}. For simplicity, we summarize here only
the main behaviour, see Eq.~(39) in Ref.\ \cite{Dermisek:2013gta},
\begin{align}\label{eq:amuformula}
  \Delta a_\mu^\text{VLL} \approx -c(m_{L,E})\frac{\lambar\lambda^E_2
    \lambda^L_2 v^3}{m_L m_E m_\mu} \times 22.5 \times 10 ^{-10}\,,
\end{align}
where in our convention the coefficient $c(m_{L,E})$ is approximately $+1$ $(-1)$ for
heavy (EW-scale) VLL masses.
Unlike in most other models where the chirally enhanced contributions to the effective Higgs coupling and $a_\mu$
arise simultaneously at the one-loop level
(cf. Ref.~\cite{Crivellin:2021rbq}), in the VLL model, new
contributions to $m_\mu$ and to $\lambda_{\mu\mu}$ are already generated at tree-level
by the lepton mixing
\cite{Kannike:2011ng,Dermisek:2013gta,Poh:2017tfo,Dermisek:2023nhe}. Hence,
there is a very peculiar correlation between the effective Higgs
coupling $\lambda_{\mu\mu}$ and $\Delta a_\mu^\text{VLL}$. For  heavy
VLL masses  $m_{L,E}\gg v$  it can be written as
\begin{align}\label{eq:tree-level-correlation}
	\bigg|\frac{\lambda_{\mu\mu}}{\lambda_{\mu\mu}^\text{SM}}\bigg|^\text{tree} \approx \quad 
	\bigg| 1 - \frac{2\Delta a_\mu^\text{VLL}}{22.5\times 10^{-10}} \bigg|.
\end{align}
This equation is the origin of Eq.~\eqref{eq:amuVLL} quoted in the
Introduction. 

\begin{figure}
	\centering
	\begin{align*}
		\begin{tikzpicture}[baseline=-.5ex, scale=.8]
			\begin{feynman}[small]
				\vertex (m1) at (-2,0) {$\mu_R$};
				\vertex (m2) at (2,0) {$\mu_L$};
				\vertex (mh1) at (-1,0);
				\vertex (mh2) at (0,0);
				\vertex (mh3) at (1,0);
				\vertex (h1) at (-1,1) {$H$};
				\vertex (h2) at (0,1) {$H$};
				\vertex (h3) at (1,1) {$H$};
				\diagram*{
					(m1) -- [fermion] (mh1) -- [fermion, thick, edge label'=$E_L$] (mh2) -- [fermion, thick, edge label'=$L_R$] (mh3) -- [fermion] (m2),
					(mh1) -- [scalar] (h1), (mh2) -- [scalar] (h2), (mh3) -- [scalar] (h3),
				};
			\end{feynman}
		\end{tikzpicture} \sim \frac{\lambda^E_2\lambar\lambda^L_2}{m_L m_E}
		\hspace{2cm}
		\begin{tikzpicture}[baseline=-.5ex, scale=.8]
			\begin{feynman}[small]
				\vertex (h) at (-1.5,0) {$H$};
				\vertex (m1) at (1.5,1) {$\mu_L$};
				\vertex (m2) at (1.5,-1) {$\mu_R$};
				\vertex (a) at (-.5,0);
				\vertex (b) at (.5,.5);
				\vertex (c) at (.5,-.5);
				\diagram*{
					(h) -- [scalar] (a) -- [fermion, thick, edge label=$L_R$] (b) -- [fermion] (m1),
					(m2) -- [fermion] (c) -- [fermion, thick, edge label=$E_L$] (a),
					(c) -- [scalar] (b)
				};
			\end{feynman}
		\end{tikzpicture} \sim \frac{\lambda^E_2\lambar\lambda^L_2}{16\pi^2}
	\end{align*}
	\caption{\emph{Left:} matching diagram for $C_{\mu H}$ at tree-level. \emph{Right:} matching diagram for $y_\mu^\text{EFT}$ at one-loop order.
		The thick lines represent the heavy internal VLL that are integrated out.}
	\label{fig:EFTdiagrams}
\end{figure}
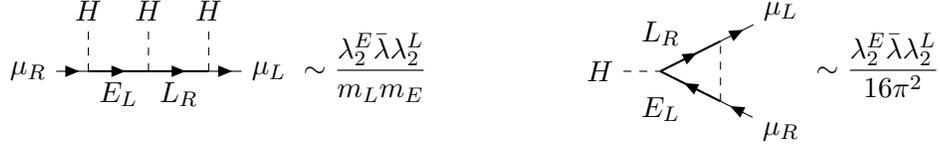

Previous studies \cite{Kannike:2011ng,Dermisek:2013gta} focused in
particular on the region of small VLL masses $m_{L,E}\sim v$ where
$c(m_{L,E})\approx-1$ and the sign in the correlation  
changes, such that $\Delta a_\mu^\text{VLL}$ of order of the Fermilab result \cite{Muong-2:2023cdq,Aoyama:2020ynm} would require
a large deviation $|\lambda_{\mu\mu}/\lambda_{\mu\mu}^\text{SM}|\sim 3$. After these studies, LHC constraints on the Higgs coupling
have become much more stringent, restricting $|\lambda_{\mu\mu}/\lambda_{\mu\mu}^\text{SM}| \simeq 1$. This has been used
as a motivation for extended VLL models, where the correlation Eq.~\eqref{eq:tree-level-correlation} is modified e.g. by
couplings to additional scalar particles \cite{Chun:2020uzw,Dermisek:2015oja,Dermisek:2020cod}. On the other hand, for large
VLL masses the Fermilab result and LHC data are perfectly compatible with Eq.~\eqref{eq:tree-level-correlation} if $\lambda_{\mu\mu}/\lambda_{\mu\mu}^\text{SM} \simeq -1$. It is therefore of interest to better understand how one-loop corrections to the effective Higgs coupling affect the above correlation.

The effective field theory (EFT) approach used in Refs.~\cite{Kannike:2011ng,Dermisek:2013gta} to study the correlation due to the lepton mixing also nicely
extends to the one-loop level. After integrating out the VLLs, 
the resulting Lagrangian contains two chirality-flipping terms, a SM-like Yukawa interaction and a dimension-6 operator with three Higgs fields,
\begin{align}
	\label{eq:Leff}
	\L_{EFT} &\supset - y^\text{EFT}_\mu \bar{l}^2_L \mu_R H
	- C_{\mu H} ~  \bar{l}^2_L \mu_R H (H^\dagger H) + h.c.\,.
\end{align}
Both $y^\text{EFT}_\mu$ and the Wilson coefficient $C_{\mu H}$ are determined by matching and admit an expansion into tree-level and 
loop contributions (see Fig.~\ref{fig:EFTdiagrams})
\begin{align}\label{eq:matchingresult}
	y_\mu^\text{EFT} = y_\mu + \O\Big(\frac{\lambar\lambda^E_2\lambda^L_2}{16\pi^2}\Big), \qquad C_{\mu H} = \frac{\lambar\lambda^E_2\lambda^L_2}{m_L m_E} \Big(1+ \O(1L) \Big).
\end{align}
Both the tree-level matching contribution to $C_{\mu H}$ as well as the one-loop matching contribution to $y_\mu^\text{EFT}$ are chirally enhanced compared
to the fundamental Yukawa coupling $y_\mu$, with the difference that the tree-level contribution is suppressed by $1/(m_L m_E)$ while the one-loop 
contribution is suppressed by $1/(16\pi^2)$.
In the EFT, the muon pole mass $m_\mu$ and the physical muon--Higgs coupling $\lambda_{\mu\mu}$ are given by
\begin{subequations}\label{eq:EFTmYuk}
	\begin{align}
		m_\mu &= y_\mu^{\text{EFT}} v + ~~ C_{\mu H} v^3  + \Delta m_\mu^{\text{1L,EFT}},\\
		\lambda_{\mu\mu} &= y_\mu^{\text{EFT}} ~~ + 3 C_{\mu H} v^2 + \Delta \lambda_{\mu\mu}^{\text{1L,EFT}}.
	\end{align}
\end{subequations}
The important factor $3$ arises from combinatorics and the last terms
correspond to loop corrections within the EFT. Those EFT loop
corrections can arise e.g.\ from loops with gauge or Higgs bosons and
cannot contain additional chiral enhancements. Chiral enhancements
only appear in the matching coefficients as discussed above.
Interestingly, for VLL masses above the TeV scale the one-loop corrections
entering via $y_\mu^{\text{EFT}}$ can be larger than the tree-level
corrections entering via $C_{\mu H}$, which provides further
motivation for a one-loop analysis.

Eliminating $y_\mu^{\text{EFT}}$ between the previous two equations,
using Eqs.~\eqref{eq:amuformula} and \eqref{eq:matchingresult} and
neglecting higher orders leads to the lowest-order correlation formula
(\ref{eq:tree-level-correlation}). Taking into account higher orders,
we find that no new chiral enhancements beyond those present at tree-level are introduced by the loop corrections
and the one-loop corrected correlation for large VLL masses
is of the form\footnote{If other dimension-6 operators besides $\bar{l}^2\mu H^3$ are included, 
	additional mass-suppressed terms can appear independently of $\Delta a_\mu^\text{VLL}$. In our case none of the possible operators introduce
	new chiral enhancements and are therefore neglected. }
\begin{align}\label{eq:oneloopcorrelation}
	\bigg| \frac{\lambda_{\mu\mu}}{\lambda_{\mu\mu}^{\text{SM}}} \bigg|^{\text{tree+1L}}
	&\approx \quad 	
	\bigg|1 - \frac{2\Delta a_\mu^{\text{VLL}}}{22.5\times10^{-10}}\times\Big(1+\O(\text{1L})\Big)\bigg|.
\end{align}
This can also be compared with the analysis of Ref. \cite{Crivellin:2021rbq}, where generic models without tree-level mixing were considered, such
that $C_{\mu H}$ arises only at the one-loop level. Going through similar steps, a correlation like Eq.~\eqref{eq:oneloopcorrelation}
can be derived, but $\Delta a_\mu$ would only appear multiplied with
$\mathcal{O}(\text{1L})$ effects (i.e.\ the ``$1+$'' term in the round
brackets would be
absent in Eq.~\eqref{eq:oneloopcorrelation}).

\begin{figure}
	\centering
	\includegraphics[width=.45\textwidth]{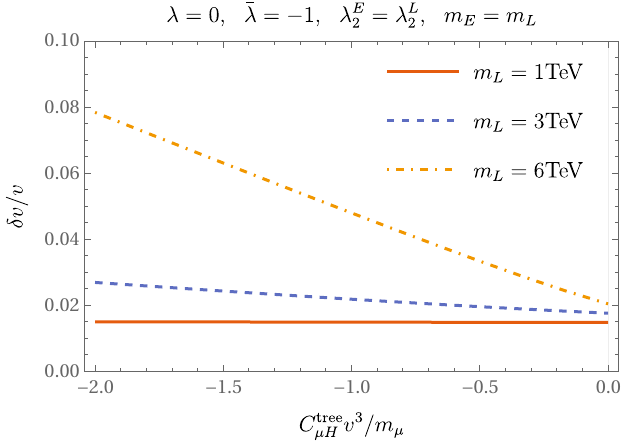}\hfill
	\includegraphics[width=.45\textwidth]{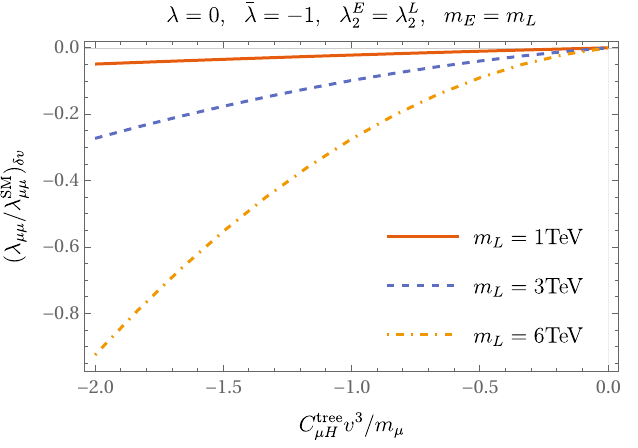}
	\caption{\emph{Left:} Plot of $\delta v/v$ as a function of $C_{\mu H}$ for several fixed values of the VLL masses.
		\emph{Right:} Plot of the corresponding contribution to $\lambda_{\mu\mu}/\lambda_{\mu\mu}^\text{SM}$.}
	\label{fig:dv-contribution}
\end{figure}

Eq.~\eqref{eq:oneloopcorrelation} defines our expectation for the
behaviour of the explicit one-loop results discussed below.
Despite the absence of new chiral enhancements at one-loop order, it
turns out that the magnitude of the one-loop corrections can be
expected to be large, due to logarithms of the VLL masses and mixing
effects. To provide a qualitative estimate, we mention that one particular contribution that exhibits both of these effects is the
correction due to the renormalization constants $\delta v$ and $\delta Z_h$. These are obtained from gauge- and Higgs-boson self-energy diagrams,
where loops with heavy fermions typically result in large
logarithmically enhanced contributions which enter the effective Higgs coupling through the 
Yukawa counterterm and field renormalization.
For example,  the leading VLL Yukawa contribution to $\delta v$ in the
limit $m_L\simeq m_E \gg v$ is given by
\begin{align}
	\frac{\delta v}{v}\Big|_\text{VLL} \simeq \frac{1}{32\pi^2} \Big(\lambda^2+\lambar^2+(\lambda^E_2)^2+(\lambda^L_2)^2\Big) \ln(\frac{m_L^2}{\bar\mu^2}) 
\end{align}
Similarly, the contribution due to the top quark reads
\begin{align}
	\frac{\delta v}{v}\Big|_{m_t} \approx \frac{3 y_t^2}{32\pi^2} \bigg[\ln(\frac{m_t^2}{\bar\mu^2}) + \frac{1-2s_W^2}{2s_W^2} \bigg].
\end{align} 
Both contributions involve potentially large coefficients and a logarithmic scale dependence, such that $\delta v/v$ will always result in large contributions to the effective Higgs coupling,
regardless of the choice of renormalization scale.

If the lepton masses were
generated purely by the Higgs mechanism as in the SM, 
only the combination $\delta Z_h/2-\delta v/v$ would appear in the
one-loop effective Higgs coupling (this combination corresponds to the
quantity studied in
Refs.\ \cite{Sperling:2013eva,Sperling:2013xqa}). In this combination
the logarithmically enhanced terms 
cancel and the remaining correction is at most $\O(1\%)$, even if heavy new particles are present. 
However, in the VLL model additional terms $\propto\delta v$ exist due to the 
tree-level lepton mixing. This is most directly seen by applying the
transformation $v\to v+\delta v$ and the Higgs field renormalization in the tree-level result,
\begin{align}
	\lambda^\text{tree}_{\mu\mu} = \tfrac{m_\mu}{v} + 2C_{\mu H} v^2 \quad \implies \quad \lambda_{\mu\mu}^\text{1L}\big|_{\delta v} 
	= \big(4 C_{\mu H} v^2 - \tfrac{m_\mu}{v}\big)\tfrac{\delta
          v}{v}
        + \lambda_{\mu\mu}^\text{tree} \tfrac{1}{2}\delta Z_h.
\end{align}
Simplifying, this results in the following contribution to the
one-loop effective muon--Higgs coupling,
\begin{align}
	\lambda^\text{1L}_{\mu\mu}\big|_{\delta v} = \lambda_{\mu\mu}^\text{tree} (\tfrac{1}{2}\delta Z_h - \tfrac{\delta v}{v}) + 6 C_{\mu H} v^2 \tfrac{\delta v}{v},
\end{align}
where the first term is small but the second term is doubly enhanced
by the prefactor $6$ and the large contributions within $\delta v$.
In particular in the region where $\lambda_{\mu\mu}/\lambda_{\mu\mu}^\text{SM}\approx -1$
and the VLL masses are $\gtrsim 1~\text{TeV}$, the contribution of $\delta v/v$ can account for several $\O(10\%)$ corrections to the tree-level value. 
This effect is illustrated in Fig. \ref{fig:dv-contribution}
where the value of $\delta v$ (left) and corresponding correction to $\lambda_{\mu\mu}$ (right) are plotted for several different VLL masses.

Including the contributions from all loop and counterterm diagrams
also beyond $\delta v$, 
the leading logarithmic one-loop VLL Yukawa contribution to the effective Higgs coupling is given by
\begin{align}\label{eq:eff-cpl-leading-term}
	\frac{\lambda_{\mu\mu}}{\lambda_{\mu\mu}^\text{SM}}\Big|_\text{1l,VLL} \simeq \frac{\lambda^E_2\lambda^L_2\lambar v^3}{32\pi^2 m_L^2 m_\mu}
	\bigg[ 9 \Big(\lambda^2 + \lambar^2 + (\lambda^E_2)^2 + (\lambda^L_2)^2\Big) - 12 \lambda\lambar\bigg] \ln(\frac{m_L^2}{M^2_h}). 
\end{align}
Note that the terms in the square brackets are always positive and thus lead to an increase of $|\lambda_{\mu\mu}|$ compared to the tree-level result.

\subsection{Numerical Results}

\begin{figure}
	\centering
	\includegraphics[width=.45\textwidth]{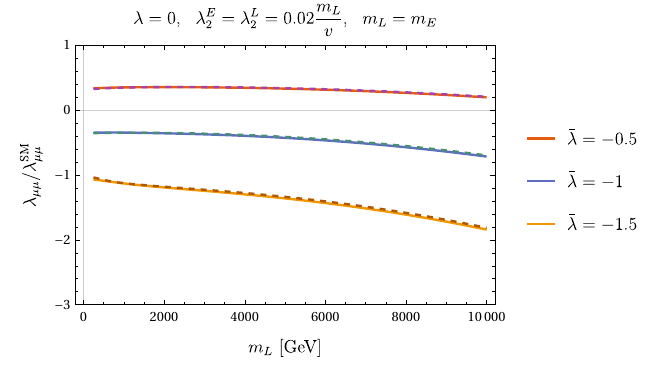}\hspace{1cm}
	\includegraphics[width=.45\textwidth]{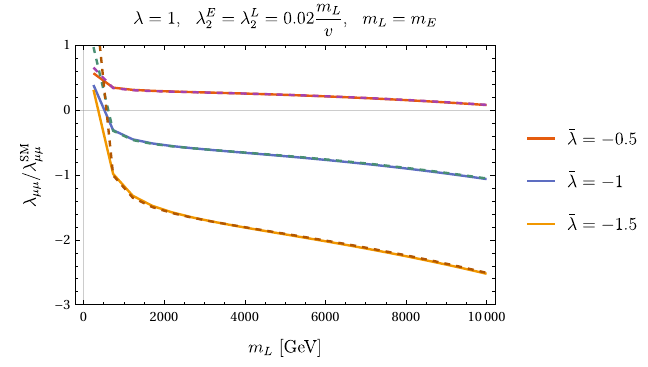}
	\caption{Comparison between the exact result (solid) and fit (dashed) of the one-loop results for $\lambda_{\mu\mu}/\lambda_{\mu\mu}^\text{SM}$ as a function of $m_L$.
		We have set $m_E = m_L$ and $\lambda^E_2=\lambda^L_2 = 0.02 \frac{m_L}{v}$.	The colours correspond to different values of $\lambar$ as indicated, while
		$\lambda = 0$ (left) and $\lambda = 1$ (right).
	}
	\label{fig:Rmu-ML}
\end{figure}

Finally  we present numerical results
on the complete one-loop corrections to $\lambda_{\mu\mu}$ in our renormalization scheme.
We find good agreement with the qualitative discussion above and determine the numerical impact on the correlation Eq.~\eqref{eq:oneloopcorrelation}
between $\lambda_{\mu\mu}$ and $\Delta a_\mu^\text{VLL}$. We note that the updated analysis of Ref.~\cite{Lourenco:2023sde} 
on the EW precision constrains stemming from Z-pole observables translates into the following bounds on $\lambda^E_2$ and $\lambda^L_2$
\begin{align}\label{eq:EW-constraints}
	\bigg|\frac{\lambda^L_2 v}{m_L}\bigg| \lesssim 0.02, \qquad \bigg(\frac{\lambda^E_2 v}{m_E}\bigg)^2  + 1.16 \bigg(\frac{\lambda^L_2 v}{m_L}\bigg)^2\lesssim 0.00054.
\end{align}
In particular, in order to achieve $\lambda_{\mu\mu}^\text{tree}=-\lambda_{\mu\mu}^\text{SM}$ these bounds require $|\lambar| \gtrsim 1.5$.\footnote{%
These constraints are potentially slightly modified by inclusion of the one-loop corrections to the muon--Z coupling which, however, 
are beyond the scope of this paper. }

\begin{figure}
	\centering
	\includegraphics[width=.45\textwidth]{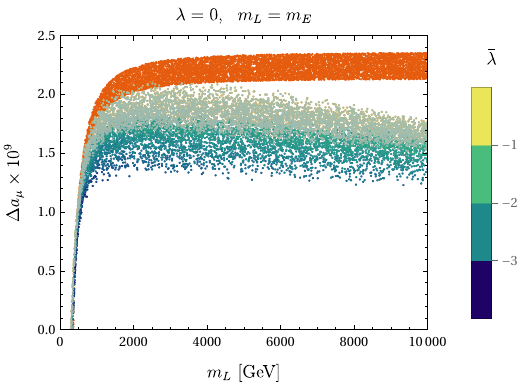}\hspace{1cm}
	\includegraphics[width=.45\textwidth]{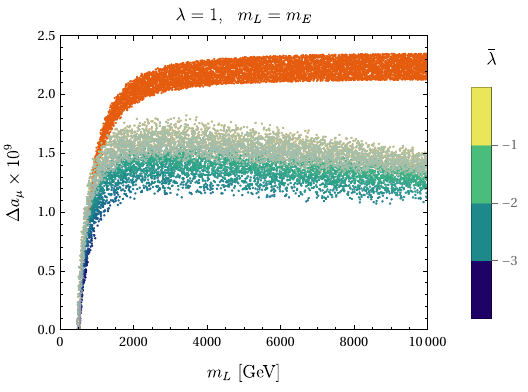}
	\caption{Scatter plot of the correlation between $\Delta a_\mu^\text{VLL}$ and $m_L$ for $\lambda_{\mu\mu}/\lambda^\text{SM}_{\mu\mu}=-1\pm 0.1$. 
		The red points show the results using the tree-level value of $\lambda_{\mu\mu}$ while the one-loop results are colour-coded to highlight the dependence on $\lambar$.
		The greyed-out points violate the EW constraints Eq.~\eqref{eq:EW-constraints}.
		The data was obtained from a random scan with $m_L=m_E\in [250,10000]$ GeV and $\lambda=0$ (left), $\lambda=1$ (right).
	}
	\label{fig:amu-ML}
\end{figure}

We have performed a grid and a random scan for $m_L = m_E \in [250,10000]$ GeV, $\lambar\in [-4,0]$, $\lambda\in [-1,1]$ and $\lambda^{L/E}_2$ in the range
allowed by the above constraints. Using the grid scan data together with an informed ansatz following the discussion in the previous section, we obtain
the following approximate numerical formula for the one-loop corrected normalized effective Higgs-coupling
\begin{align}\label{eq:fit}
	\begin{split}
		\frac{\lambda_{\mu\mu}}{\lambda^{SM}_{\mu\mu}} \approx 1 + \frac{\lambda^E_2\lambda^L_2\lambar v^3}{m_L^2 m_\mu}&\bigg\{ 2+0.056 - 0.032\ln(\tfrac{m^2_L}{M^2_h}) + \frac{v^2}{m_L^2}\Big[4\lambar\lambda
		- (\lambda^E_2)^2 - (\lambda^L_2)^2\Big] \\
		&  + 0.028 \Big[\lambar^2+\lambda^2+(\lambda^L_2)^2+(\lambda^E_2)^2\Big] \ln(\tfrac{m_L^2}{M_h^2}) - 0.038 \lambda\lambar \ln(\tfrac{m_L^2}{M_h^2}) \\
		&  - 0.035 \Big[\lambar^2+\lambda^2+(\lambda^L_2)^2\Big] + 0.033 (\lambda^E_2)^2 + 0.013 \lambda\lambar\bigg\}.
	\end{split}
\end{align}
The ``$2$'' in the brackets corresponds to lowest order, all other
terms are
one-loop corrections. The logarithmically enhanced terms correspond to
the analytical result (\ref{eq:eff-cpl-leading-term}). In the fit we have also
included $\O(v^2/m_L^2)$ corrections corresponding to the tree-level expansion of $Y^h_{22}$ 
to improve the convergence at small VLL masses.
In Fig.~\ref{fig:Rmu-ML} we have plotted a comparison between the numerical fit (dashed) and the results obtained from the full loop calculation and 
tree-level couplings from the exact diagonalization (solid) as a
function of $m_L$. The plot shows good agreement at large VLL masses  $\gtrsim 1000$ GeV, but
a deviation at small masses (in particular for $\lambda\neq 0$) which
can be attributed to the missing terms of even higher orders in $v/m_L$.

Finally, in Fig.~\ref{fig:amu-ML} we have plotted points from the random scan in the $m_L$--$\Delta a_\mu^\text{VLL}$ plane, for which
$\lambda_{\mu\mu}/\lambda_{\mu\mu}^\text{SM}=-1\pm 0.1$.
The red points show the results using $\lambda_{\mu\mu}^\text{tree}$ and thus reflect the correlation Eq.~\eqref{eq:tree-level-correlation}
leading to the asymptotic value $\Delta a_\mu^\text{VLL} \approx 22.5\times 10^{-10}$.
In contrast, including the one-loop results (color-coded according to the value of $\lambar$) leads to a significantly lower asymptotic value
of $\Delta a_\mu^\text{VLL}$. This can be understood as a consequence of the logarithmic enhancement of the one-loop corrections. 
According to Eq.~\eqref{eq:fit}, in combination the logarithmic terms can easily increase the effective Higgs coupling by $10$--$50\%$ compared to the tree-level value,
which in turn requires smaller couplings to fulfil $\lambda_{\mu\mu}/\lambda_{\mu\mu}^\text{SM}\approx -1$ at one-loop order.
Consequently, with the asymptotic behavior of $\Delta a_\mu^\text{VLL}$ determined by Eq.~\eqref{eq:amuformula}, the one-loop corrections to the 
correlation result in an overall decrease of the predicted value of the deviation.

\section{Conclusion}\label{sec:conclusion}

In models with vector-like leptons, the lepton mass generation
mechanism can be strongly modified. Similar to the neutrino seesaw
mechanism, SM-like lepton masses can receive significant contributions
from mixing with VLL with fundamental gauge-invariant Dirac mass terms. The
modified mass generation mechanism is accompanied by potentially large effects
in the effective lepton--Higgs couplings and in the anomalous magnetic
moment $a_\mu$, which are strongly correlated in a peculiar way.

Here we have started an analysis of higher-order effects in a VLL
model where one VLL generation can mix with all SM leptons via additional
Yukawa couplings. As the first theoretical step we have developed a
renormalization scheme. Our scheme addresses the mixing between the SM
and vector-like leptons and implements full on-shell conditions on the 
lepton self-energies. It thus establishes a direct physical
interpretation of mass parameters and fields and allows a transparent
computation of observables. The remaining freedom is fixed by imposing
\MSbar conditions on a judiciously chosen set of fundamental
parameters of the gauge-eigenstate Lagrangian.
Here we use the redundancy Eq.~\eqref{eq:wlog-transformation} between
parameters and fields to minimize the number of such parameters.

Because of the combination of on-shell and \MSbar conditions, there is
a residual renormalization-scale dependence
of the fundamental model parameters. The corresponding $\beta$
functions differ from standard ones because of the finite parts of
counterterms and because of the use of the redundancy. Both changes
have been evaluated. The cancellation of divergences as well as the
cancellation of the renormalization-scale dependence in the
computation of observables provides a strong check of the scheme and
its implementation.

As the first phenomenological application we have calculated the
one-loop corrections to the effective muon-Higgs coupling
$\lambda_{\mu\mu}$ and its correlation with 
the VLL contribution to the muon magnetic moment $\Delta
a_\mu^\text{VLL}$. We focused in particular on the interesting 
region $\lambda_{\mu\mu}\simeq - \lambda_{\mu\mu}^\text{SM}$, where
the tree-level correlation predicts $\Delta a_\mu^\text{VLL}$ roughly  
of order of the current tentative experimental deviation. We found
that, despite not introducing 
new chiral enhancements, the one-loop corrections can significantly
modify the results found at tree-level and in particular lead to a
smaller predicted value  
of $\Delta a_\mu^\text{VLL}$. This is mainly due to the logarithmic
enhancements present in the one-loop results for $\lambda_{\mu\mu}$,  
together with the large couplings required to achieve
$\lambda_{\mu\mu}\simeq - \lambda_{\mu\mu}^\text{SM}$.

Quantitatively, the lowest-order correlation Eq.~\eqref{eq:amuVLL}
predicts a rigidly fixed, large value for  $\Delta a_\mu^\text{VLL}$
of around $22.5\times10^{-10}$,
but the loop-corrected correlation is incompatible with such large
values of $\Delta a_\mu^\text{VLL}$ and predicts instead a range
between around $(10\ldots18)\times10^{-10}$ for masses in the
multi-TeV region. It will be interesting to compare this range to
future updates of the Fermilab $a_\mu$ measurement and the SM theory
prediction.

As a further outlook, given the large corrections found here
it will be motivated to explore the phenomenological impact of
one-loop corrections and the properties of the renormalization scheme
in a broader range of observables. The renormalization scheme can be
used to 
evaluate further observables in the lepton/Higgs/EW sector at the loop
level, for example to obtain more accurate EW constraints on the model
from Z-pole measurements. Moreover, the scheme is not restricted to
the flavour-conserving case but also allows to compute e.g.\ lepton-flavour
violating Higgs-, Z-, or muon decays.

\section*{Acknowledgments}
 We acknowledge financial support by the German Research Foundation (DFG) under grant number STO 876/7-2.

\appendix
\section{Feynman Rules} \label{sec:Feynman-Rules}
Here we list Feynman rules of the VLL model in the mass-eigenstate
basis defined in Eq.~\eqref{eq:mass-basis}. The SM vertices are taken
from Ref.\ \cite{Romao:2012pq}.
For the gauge sector we adapt the sign convention
from \cite{Denner-Böhm:2001} who define the gauge-boson mass
eigenstates by 
\begin{align}
	\begin{pmatrix}	~W^+_\mu~ \\ W^-_\mu \end{pmatrix} = 
	\begin{pmatrix}	\frac{1}{\sqrt{2}} & -\frac{i}{\sqrt{2}} \\ \frac{1}{\sqrt{2}} & \hphantom{-}\frac{i}{\sqrt{2}} \end{pmatrix}
	\begin{pmatrix}	~W^1_\mu~ \\ W^2_\mu \end{pmatrix}, \qquad 
	\begin{pmatrix}	~A_\mu~ \\ Z_\mu  \end{pmatrix} = 
	\begin{pmatrix}	c_W & - s_W \\ s_W & c_W \end{pmatrix}
	\begin{pmatrix}	~B_\mu~ \\ W^3_\mu \end{pmatrix},
\end{align}
and the covariant derivative as
\begin{align}
	D_\mu = \partial_\mu - i\tfrac{g_W}{\sqrt{2}}\left(t^+ W^+_\mu + t^- W^-_\mu \right) - i \tfrac{g_W}{c_W} (t^3 - Q s_W^2) Z_\mu + ie Q A_\mu,
\end{align}
where $t^\pm=(t^1\pm i t^2)$ (for doublets, $t^i=\frac{\sigma^i}{2}$), $Q=t^3 + Y$ and $e=g_Y c_W=g_W s_W$.
\subsection*{Gauge Interactions}
\begin{figure}[h!]
	\centering
	\begin{subfigure}[c]{.43\textwidth}
		\begin{equation*}
			\begin{tikzpicture}[baseline=-.1em, scale=.5]
				\begin{feynman}[small]
					\vertex (a) at (0,0);
					\vertex (q) at (1.75, 1) {$f_a$};
					\vertex (v1) at (-1.75,0) {$V_\mu$};
					\vertex (v2) at (1.75,-1) {$\bar f_b$};
					\diagram* {
						(v1) -- [photon] (a) -- [anti fermion] (v2),
						(a) -- [fermion] (q)
					};
				\end{feynman}
			\end{tikzpicture} = i\gamma_\mu \Big(C_R \PR + C_L \PL\Big)_{ab}
		\end{equation*}
	\end{subfigure}\hfill
	\begin{subfigure}[c]{.56\textwidth}
		\renewcommand{\arraystretch}{1.5}%
		\begin{tabular}{|c||c|c|c|c|c|}
			\hline
			$V\bar f f$ & $A \, \bar{\hat e}^b \hat e^a$ & $Z\, \bar{\hat e}^b \hat e^a$ & $Z \, \bar{\hat\nu}^b \hat \nu^a$ & $W\bar{\hat e}^b \hat \nu^a$ & $W\bar{\hat\nu}^b \hat e^a$  \\ \hline
			$C^V_R$ & $e\delta_{ab}$ & $(g^Z_R)_{ab}$ & $(g^{Z\nu}_R)_{ab}$ & $(g^W_R)_{ab}$ & $(g^W_R)^*_{ba}$ \\ \hline
			$C^V_L$ & $e\delta_{ab}$ & $(g^Z_L)_{ab}$ & $(g^{Z\nu}_L)_{ab}$ & $(g^W_L)_{ab}$ & $(g^W_L)^*_{ba}$ \\ \hline
		\end{tabular}
	\end{subfigure}\hfill
	\caption{Feynman rules for gauge interactions}
\end{figure}
where
\begin{subequations}
	\begin{align}
		g^Z_{L/R} &= \tfrac{e}{c_W s_W} U_{L/R}^\dagger \cdot \big( s_W^2 - \tfrac{1}{2}T_{L/R}\big) \cdot U_{L/R}, \\
		g^{Z\nu}_{L/R} &= \tfrac{e}{c_W s_W} U^{\nu\dagger}_{L/R} \cdot \big(\tfrac{1}{2} T_{L/R}\big) \cdot U^{\nu}_{L/R} \\
		g^W_{L/R} &= \tfrac{e}{\sqrt{2} s_W} U^{\nu\dagger}_{L/R} \cdot \big(T_{L/R}\big) \cdot U_{L/R}
	\end{align}
\end{subequations}
with $T_R = \text{diag}(0,0,0,0,1)$ and $T_L = \text{diag}(1,1,1,1,0)$.

\subsection*{Yukawa Interactions}

\begin{figure}[h!]
	\centering
	\begin{subfigure}{.43\textwidth}
		\centering
		\begin{equation*}
			\begin{tikzpicture}[baseline=-.1em, scale=.5]
				\begin{feynman}[small]
					\vertex (a) at (0,0);
					\vertex (q) at (1.75, 1) {$f_a$};
					\vertex (v1) at (-1.75,0) {$S$};
					\vertex (v2) at (1.75,-1) {$\bar f_b$};
					\diagram* {
						(v1) -- [scalar] (a) -- [anti fermion] (v2),
						(a) -- [fermion] (q)
					};
				\end{feynman}
			\end{tikzpicture} = -\tfrac{i}{\sqrt{2}}\Big(C^S_R \PR + C^S_L \PL\Big)_{ab}
		\end{equation*}
	\end{subfigure}\hfill
	\begin{subfigure}[c]{.56\textwidth}
		\centering
		\renewcommand{\arraystretch}{1.5}%
		\begin{tabular}{|c||c|c|c|c|}
			\hline
			$S\bar f f$ & $h \, \bar{\hat e}^b \hat e^a$ & $\phi^0\, \bar{\hat e}^b \hat e^a$ & $\phi^+\bar{\hat e}^b \hat \nu^a$ & $\phi^- \bar{\hat\nu}^b \hat e^a$  \\ \hline
			$C^S_R$ & $Y^h_{ab}$ & $i Y^{\phi^0}_{ab}$ & $\sqrt{2} Y^{\phi^+}_{ab}$ & $\sqrt{2} Y^{\phi^-}_{ab}$ \\ \hline
			$C^S_L$ & $Y^{h*}_{ba}$ & $-i Y^{\phi^0*}_{ba}$ & $\sqrt{2} Y^{\phi^- *}_{ba}$ & $\sqrt{2} Y^{\phi^+*}_{ba}$ \\ \hline
		\end{tabular}
	\end{subfigure}
	\caption{Feynman rules for Yukawa interactions}
\end{figure}
where
\begin{subequations}\label{eq:Yukawa-matrices}
	\begin{alignat}{4}
		&Y^h &&= U_L^\dagger && \begin{pmatrix} \, Y_{ab} & 0~ \\ 0 & \lambar~ \end{pmatrix} U_R, \\[.2cm]
		&Y^{\phi^0} &&= U_L^\dagger &&\begin{pmatrix} \, Y_{ab} & ~0 \\ 0 & -\bar\lambda~\end{pmatrix} U_R, \\[.2cm]
		&Y^{\phi^+} &&= U^{\nu\dagger}_L && \begin{pmatrix} \, Y_{ab} & 0~ \\ 0 & 0~\end{pmatrix} U_R, \\[.2cm]
		&Y^{\phi^-} &&= U_L^\dagger &&\begin{pmatrix} ~0 & ~0~ \\ ~0 & \lambar \end{pmatrix} U_R^\nu
	\end{alignat}
\end{subequations}

\section{Renormalization Group Equations}\label{App:RGE}
Here we list the one-loop $\beta$-functions of the model assuming the
off-diagonal counterterms $\delta M_L^i, \delta M_E^i$ and $\delta
y_{ij}$ have been absorbed into $\delta V_L$ and $\delta V_R$, see
Eq.\ (\ref{eq:wlog-ct}).
In general, if we consider a theory with a set of bare
parameters $\Theta^0_\alpha = \mu^{\xi_\alpha \epsilon} (\Theta_\alpha + \delta \Theta_\alpha)$,
where the $\mu$-dependent prefactor is introduced to generate a
$D$-dimensional bare Lagrangian while keeping parameters integer-dimensional, the $\beta$-functions in the
$\overline{\text{MS}}$-scheme are given by
\begin{align}
	\beta(\Theta_\alpha) = - \xi_\alpha \delta \Theta^{[1]}_\alpha + \sum_\beta \frac{\partial \delta\Theta_\alpha^{[1]}}{\partial \Theta_\beta} \xi_\beta \Theta_\beta
\end{align}
where $\delta\Theta_\alpha^{[1]}$ denotes the coefficient of the
$1/\epsilon$ pole in $\delta\Theta_\alpha$. Using this relationship
and using the abbreviations $g_W = e/s_W$, $g_Y = e/c_W$ and
\begin{align}
	\delta\Upsilon = 2\lambar^2 + 2\lambda^2 + 2 \sum_j \Big((\lambda_j^E)^2 + (\lambda_j^L)^2 + y_j^2 + 3 (y_j^u)^2 + 3 (y_j^d)^2\Big) - \frac{9}{2} g_W^2 - \frac{15}{2} g_Y^2,
\end{align}
the Yukawa $\beta$-functions are given by
\begin{subequations}\label{eq:Yukawa-beta-functions}
	\begin{align}
		&(32\pi^2)\,\beta(y_i) = ~y_i \, \bigg[\delta\Upsilon + 3 y_i^2 + 2 (\lambda^L_i)^2 + (\lambda^E_i)^2 \bigg] - 4\lambda^E_i\lambda^L_i\lambar \tfrac{m_E^2+2m_L^2}{m_E m_L} \\[.2cm]
		\begin{split}
			&(32\pi^2)\beta(\lambda^L_i) = \lambda^L_i \bigg[ d\Upsilon + 3 (\lambda^L_i)^2 + 4 y_i^2 + \lambda \Big(\lambda - \lambar \tfrac{8 m_L}{m_E}\Big)\bigg] 
			+ 2\lambda^E_i y_i \bigg[\lambda + \lambar\tfrac{2 m_E}{m_L}\bigg] \\
			&\qquad + \sum_{k\neq i} \tfrac{\lambda^L_k}{y_i^2-y_k^2} \bigg[ \lambda^L_i \lambda^L_k \Big(5y_i^2-y_k^2\Big) + 2\lambda^E_i \lambda^E_k y_i y_k 
			- 4\lambar\Big(\lambda^L_i \lambda^E_k y_k + \lambda^E_i \lambda^L_k y_i\Big) \tfrac{m_E^2+2m_L^2}{m_E m_L} \bigg] 
		\end{split} \\[.2cm]
		\begin{split}
			&(32\pi^2)\beta(\lambda^E_i) = \lambda^E_i \bigg[ d\Upsilon + 3 (\lambda^E_i)^2 + 5 y_i^2 + 2\lambda \Big(\lambda - \lambar \tfrac{2 m_E}{m_L}\Big)\bigg] 
			+ 4\lambda^L_i y_i \bigg[\lambda + \lambar\tfrac{2 m_L}{m_E}\bigg] \\
			&\qquad + \sum_{k\neq i} \tfrac{\lambda^E_k}{y_i^2-y_k^2} \bigg[ \lambda^E_i\lambda^E_k\Big(4y_i^2 - 2 y_k^2\Big) + 4 \lambda^L_i \lambda^L_k y_i y_k
			- 4 \lambar\Big(\lambda^L_i \lambda^E_k y_i + \lambda^E_i \lambda^L_k y_k \Big) \tfrac{m_E^2+2m_L^2}{m_E m_L} \bigg]
		\end{split}	\\[.2cm]
		\begin{split}
			&(32\pi^2)~\beta(\lambda) = \lambda ~ \bigg[ \delta\Upsilon + 3\lambda^2 + \sum_i\Big(4 (\lambda^E_i)^2 +5 (\lambda^L_i)^2\Big) \bigg] + \sum_i \bigg[6 y_i \lambda^E_i \lambda^L_i 
			+ 4\lambar \tfrac{m_E^2 (\lambda^E_i)^2 + 2 m_L^2 (\lambda^L_i)^2 }{m_E m_L}\bigg]
		\end{split} \\[.2cm]
		&(32\pi^2)~\beta(\lambar) = \lambar ~ \bigg[\delta\Upsilon + 3 \lambar^2 \bigg].
	\end{align}
\end{subequations}
Similarly, the $\beta$-functions of the Dirac masses are given by
\begin{subequations}\label{eq:Dirac-beta-functions}
	\begin{align}
		(32\pi^2)\beta(m_L) &= 4 \lambar\lambda m_E + \Big(\lambda^2 + \lambar^2 + \sum_j (\lambda^L_j)^2 - 9g_W^2 - 3 g_Y^2 \Big) m_L \\
		(16\pi^2)\beta(m_E) &= 4 \lambar\lambda m_L + \Big(\lambda^2 + \lambar^2 + \sum_j (\lambda^E_j)^2 - 6g_Y^2 \Big) m_E.
	\end{align} 
\end{subequations}

\printbibliography
	
\end{document}